\documentclass[onecolumn]{autart}
\usepackage{graphicx}
\usepackage{amsmath}
\usepackage{url}
\usepackage{array,tabularx}
\usepackage{graphics,color}
\usepackage{epsfig}
\usepackage{mathptmx}
\usepackage{times}
\usepackage{amssymb}
\usepackage{xfrac}
\usepackage{comment}
\usepackage{epstopdf}
\usepackage{gensymb}
\usepackage[export]{adjustbox}

\makeatletter
\DeclareRobustCommand{\qed}{%
  \ifmmode 
  \else \leavevmode\unskip\penalty9999 \hbox{}\nobreak\hfill
  \fi
  \quad\hbox{\qedsymbol}}
\newcommand{\openbox}{\leavevmode
  \hbox to.77778em{%
  \hfil\vrule
  \vbox to.675em{\hrule width.6em\vfil\hrule}%
  \vrule\hfil}}
\newcommand{\qedsymbol}{\openbox}

\newcommand{\proofname}{Proof}
\makeatother

\begin{document}

\begin{frontmatter}

\title{Real-Time Steering of Curved Sound Beams in a Feedback-based Topological Acoustic Metamaterial} 

\author[Jerusalem,TelMech]{Lea Sirota}\ead{leabeilkin@mail.tau.ac.il}, \:   
\author[Jerusalem]{Daniel Sabsovich}, \:
\author[Jerusalem,Living]{Yoav Lahini},   \:
\author[Jerusalem]{Roni Ilan} \:  and \:
\author[TelMech,Sackler,LosAlamos,Living]{Yair Shokef}
            
\address[Jerusalem]{Raymond and Beverly Sackler School of Physics and Astronomy, Tel-Aviv University, Tel-Aviv 69978, Israel}             
\address[TelMech]{School of Mechanical Engineering, Tel Aviv University, Tel-Aviv 69978, Israel}
\address[Sackler]{Sackler Center for Computational Molecular and Materials Science, Tel Aviv University, Tel-Aviv 69978, Israel}
\address[LosAlamos]{Kavli Institute for Theoretical Physics, University of California, Santa Barbara, California 93106, USA}        \address[Living]{The Center for Physics and Chemistry of Living Systems, Tel Aviv University, Tel Aviv 69978, Israel}

\begin{abstract}  

We present the concept of a feedback-based topological acoustic metamaterial as a tool for realizing autonomous and active guiding of sound beams along arbitrary curved paths in free two-dimensional space. The metamaterial building blocks are acoustic transducers, embedded in a slab waveguide. The transducers generate a desired dispersion profile in closed-loop by processing real-time pressure field measurements through preprogrammed controllers. In particular, the metamaterial can be programmed to exhibit analogies of quantum topological wave phenomena, which enables unconventional and exceptionally robust sound beam guiding. As an example, we realize the quantum valley Hall effect by creating, using a collocated pressure feedback, an alternating acoustic impedance pattern across the waveguide. The pattern is traversed by artificial trajectories of different shapes, which are reconfigurable in real-time. Due to topological protection, the sound waves between the plates remain localized on the trajectories, and do not back-scatter by the sharp corners or imperfections in the design. The feedback-based design can be used to realize arbitrary physical interactions in the metamaterial, including non-local, nonlinear, time-dependent, or non-reciprocal couplings, paving the way to new unconventional acoustic wave guiding on the same reprogrammable platform. We then present a non-collocated control algorithm, which mimics another quantum effect, rendering the sound beams uni-directional.

\end{abstract}

\end{frontmatter}

\section{Introduction}    \label{Intro}

Controlling wave propagation in acoustic systems is an essential requirement in advanced engineering applications, such as acoustic imaging, acoustic signature cloaking, noise cancellation, vibration suppression, and more. 
The idea to control sound waves by artificially designing the medium in which they propagate received a considerable interest over the years, and has recently manifested itself through the emergent concept of metamaterials. \\
Metamaterials are artificially designed structures, usually of periodic nature, composed of sub-components denoted by unit cells. 
For sufficiently large wavelengths, much larger than the lattice features, metamaterials effectively act as a continuous material, whose properties are determined by the collective dynamic behavior of their unit cells. 
As such, metamaterials can exhibit properties that are unavailable in natural materials. This capability has drawn immense attention of the scientific and engineering communities. 
The use of metamaterials in the control of wave propagation began with photonic crystals,
demonstrating negative refraction \cite{shelby2001experimental,cubukcu2003negative}, superlensing \cite{pendry2000negative}, cloaking \cite{schurig2006metamaterial,ergin2010three}, and more \cite{soukoulis2011past}. %
At a later stage the metamaterial concept was extended to acoustic and elastic systems
\cite{khelif2016phononic,craster2012acoustic}. 
Notable applications are acoustic cloaking
\cite{cummer2007one}, metamaterials with a negative acoustic refractive index that can bend, focus and shape sound fields in unconventional fashions \cite{seo2012acoustic,dubois2017observation},
acoustic leaky wave antennas \cite{rohde2015experimental}, subwavelength imaging \cite{zhu2011holey}, and many more \cite{cummer2016controlling}. 
Similarly, elastic metamaterials are used to control vibrations and waves in solid materials \cite{liu2011elastic,hu146metamaterial,chen2020elastic,sirota2019tunable,sirota2020modeling}. \\
A special class of systems that has emerged in the last few years is topological metamaterials, which draws inspiration from the condensed matter branch of quantum physics \cite{thouless1982quantized,haldane1988model,kane2005quantum,bernevig2006quantum}. 
In quantum systems, the topological properties of the electronic band-structure of solids
can be exploited to achieve unique and exciting functionalities. 
One such functionality, known as topological wave phenomena, is electrical insulation in the solid interior, while conduction of current is supported only along edges, interfaces or boundaries. Remarkably, these edge waves are immune to backscattering in the presence of a broad class of imperfections and impurities, including localized defects and sharp corners. 
The role of topology manifests itself in the ability to predict the boundary properties
of finite sized materials only by
knowing the bulk properties of infinite sized materials \cite{franz2013topological}. 
The robustness of the boundary wave properties, captured by topological protection, and the exceptional immunity of the waves to back-scattering
has recently inspired the search for analogies in classical systems, substituting the electronic band-structure with acoustic or photonic dispersion relations. 
Generating topological waves in acoustics or elasticity is particularly advantageous, due to the ability to shape these waves beam-like narrow, which is obviously uncommon for sound or vibration. \\
As a result, metamaterials supporting topologically protected wave propagation have been realized in diverse fields, including photonics \cite{lu2014topological,rechtsman2013photonic}, optomechanics \cite{peano2016topological}, acoustics \cite{yang2015topological,zhang2017topological,yves2017topological}, elasticity \cite{sussman2016topological,pal2017edge,chaunsali2018subwavelength}, and more. 
While there are several classes of quantum topological effects, each having a different underlying physical mechanism \cite{franz2013topological}, the common requirement in their realization is breaking a certain form of symmetry of the system.
One class employs breaking time reversal symmetry and results in uni-directional edge waves \cite{haldane1988model,wang2015topological,nash2015topological}. 
Another class is achieved by breaking spatial symmetry in a periodic lattice (while preserving time reversal symmetry), which supports bi-directional edge waves \cite{kane2005quantum,bernevig2006quantum}. 
The Quantum Valley Hall Effect (QVHE) \cite{pan2014valley}, which we employ in our work as a representative example, belongs to this class, and can be realized in a structure as simple as a bipartite lattice with a single degree of freedom per site.
Attaching two such lattices with flipped partitions will support an exceptionally robust wave propagation along the interface.
Examples of mechanical and acoustic topological metamaterials that invoke spatial symmetry breaking, include altering the spacing between scatterers \cite{zhang2017topological} or bottle-like Helmholtz resonators \cite{yves2017topological} in acoustic waveguides, shifting elastic resonators on plates \cite{chaunsali2018subwavelength}, modifying spring constants in mass-spring lattices \cite{zhou2018quantum}, or designing arrays of pendula with intricate couplings \cite{susstrunk2015observation}. Particularly, the QVHE was demonstrated in a vibrating plate with elastic resonators of two different masses \cite{pal2017edge}, in an acoustic lattice with scatterers of two different refractive indices \cite{zhang2018achieving}, or in a flexible membrane sprayed by rigid particles of two different radii \cite{zhou2020voltage}. \\
To date, most of the metamaterial design is based on fixed elements, where the unit cells have given shape and dimensions. Such designs result in fixed dynamic properties, including effective constitutive parameters, interactions between sites, dispersion relation, etc., which are also limited to a particular operating frequency.
For a topological metamaterial, such a design would result, for example, in a single quantum effect being mimicked, with a single waveguiding trajectory at a fixed frequency range.
These limitations inspired recent attempts to construct topological metamaterials with tunable properties
\cite{darabi2020experimental,hofmann2019chiral,scheibner2020non,rosa2020dynamics,brandenbourger2019non,kotwal2019active,sirota2020feedback,sirota2020feedbackA}, where active elements were incorporated in artificial mass-spring lattices, elastic sheets and electric circuits. \\
In this work we present a method to convert a bare slab into an autonomous topological acoustic waveguide. 
The topological properties are created exclusively by real-time feedback operation of embedded acoustic transducers, and can be tuned and reconfigured by changing a control program alone, without any structural modifications.
{In Sec. \ref{Setup} we describe in detail the acoustic platform for the autonomous metamaterial, assuming a general feedback correlation between acoustic sensors and actuators. In Sec. \ref{QVHE} we derive control algorithms that realize acoustic analogues of two different quantum topological wave phenomena on this platform in real-time. 
In Sec. \ref{Time_sim} we demonstrate, using dynamical simulations, the guiding of robust and back-scattering-immune curve acoustic beams in real-time, in different frequency ranges, along reprogrammable curved trajectories. The work is discussed and summarized in Sec. \ref{Summary}.}

\section{Feedback-based acoustic metamaterial setup}   \label{Setup}

\begin{figure}[bp]
\centering
\includegraphics[height=5.0cm]{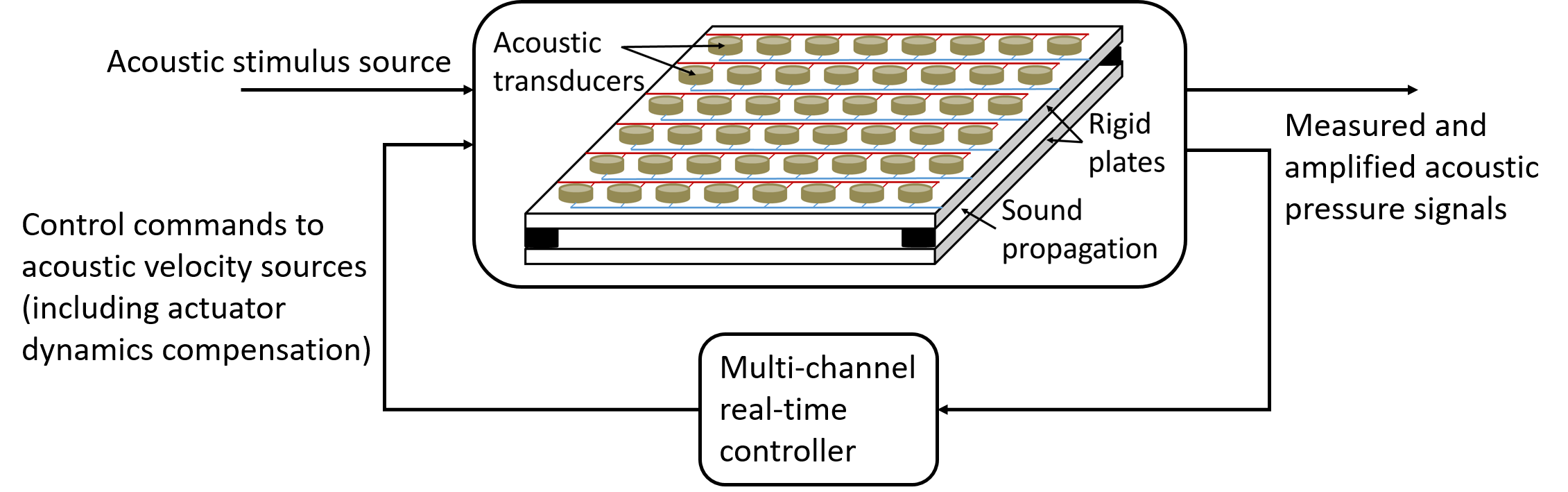}
\caption{{Feedback-based topological acoustic metamaterial setup - the physical platform.}}
\label{GenScheme}
\end{figure}

The general principle of a feedback-based metamaterial design is that the couplings between the metamaterial sites, and its consequent dynamical properties, are determined by a reprogrammable electronic feedback controller \cite{sirota2020feedback}.
The underlying mechanism includes application of external inputs to a host structure, which depend on measured responses in selected locations. 
The measurements are processed and fed back in real-time according to targeted closed-loop schemes. 
The control actuators, which are transducers embedded in a base waveguide, constitute the metamaterial unit cells. 
The particular transducers depend on the implementation platform. 
In this work we focus on an acoustic platform. \\
We consider a model for an acoustic waveguide, consisting of two rigid parallel plates separated by a small air gap, as illustrated in Fig. \ref{GenScheme}. 
The waveguide supports a continuous two-dimensional sound propagation between the plates. 
The upper plate is hollowed in a periodic pattern, embedding an array of identical acoustic actuators (loudspeakers), facing the air gap through the holes. 
The pattern defines a discrete lattice on top of the continuous acoustic field, with the actuators constituting the lattice sites. 
The choice of the holes spacing determines the lattice constant, and sets a limit for the actuators external diameter. 
The space between the plates along the metamaterial edges can be either sealed or left open, depending on the desired boundary conditions. 
The actuators can be regarded either as a source of flow velocity or of pressure \cite{pierce1990acoustics,ginsberg2018acoustics,curtain2009transfer}. In this work we regard them as acoustic velocity source generators, the role of which is creating a desired acoustic pressure field between the plates. 
The pressure field is measured by acoustic sensors (microphones) that are embedded in the waveguide along the same pattern as the actuators. 
The sensors can be either attached to the actuators themselves or mounted in mirror positions on the opposite plate, facing inwards, and assumed small enough to not significantly disturb the measured field. 
{The measured signals are processed by synchronized micro-processors, denoted by a multi-channel controller in Fig. \ref{GenScheme}, according to a pre-programmed algorithm, and fed back to the actuators. The algorithm might include pre-cancellation of the actuator self dynamics \cite{becker2018immersive}.}  %
This real-time closed-loop operation constitutes the underlying mechanism of our feedback-based acoustic metamaterial. 
The processing algorithm, as well as the exact mapping from the sensors to the actuators, depends on the particular couplings that need to be created, and is exclusively defined by the control program. 
For a wavelength large enough compared to the distance between the waveguide plates, $d$, the propagation of acoustic pressure field $p$ between the plates may be considered in the $x-y$ plane only. 
We model the coupling of this pressure field to acoustic velocity inputs $v_j$, generated by actuators at arbitrary locations $\textbf{R}_j$ in the upper waveguide plate, as
\begin{equation}  \label{eq:res_2D_OL}
c^2\nabla^2p(\textbf{r};s)=s^2p(\textbf{r};s)-\eta\rho c^2 s\sum_{\textbf{R}_j} v_j(s)\delta(\textbf{r}-\textbf{R}_j).
\end{equation}
Here $c=340$ $[m/s]$ is the speed of sound in air, $\rho=1.21$ $[kg/m^3]$ is the mass density of air, $\delta(\textbf{r}-\textbf{R}_j)$ $[1/m^2]$ is the Dirac delta function indicating the location $\textbf{R}_j$ within a unit-cell, and $\eta=A_n/d$ $[m]$, where $A_n$ is the area of the actuator active surface. 
$s$ is a complex variable in Laplace domain, which is commonly used in control design, and is related to the frequency domain as $s=i\omega$.
{The control law for the velocity inputs has the general form
\begin{equation}  \label{eq:Cont_law}
   \left[\begin{array}{c} v_1(s) \\ v_2(s) \\ \vdots \\ v_{N_a}(s) \end{array}\right]=\left[\begin{array}{c c c c}
H_{1,1}(s) & H_{1,2}(s) & ... & H_{1,{N_a}}(s) \\
H_{2,1}(s) & H_{2,2}(s) & ... & H_{2,{N_a}}(s) \\
... & ... & ... & ... \\
H_{{N_a},1}(s) & H_{{N_a},2}(s) & ... & H_{{N_a},{N_a}}(s)
\end{array}\right]\left[\begin{array}{c}p(\textbf{R}_1;s) \\ p(\textbf{R}_2;s) \\ \vdots \\ p(\textbf{R}_{N_a};s)\end{array}\right],
\end{equation}
where $N_a$ is the total number of actuators, and $H_{i,j}(s)$ are the controllers to be designed.
This control scheme can generate a metamaterial with any desired couplings (within the hardware limits) between the sites, including non-collocated interactions. The closed-loop stability must be verified for each particular design.}
When control is turned off and the actuators are assumed ideal, i.e. their surface is rigid when inactive, the target system comprises fully sealed cavities, retaining the slab waveguide. 
In the next section we derive the controllers in \eqref{eq:Cont_law} to realize an acoustic analogue of the QVHE.

\section{{Control algorithms generating topologically protected wave propagation}}   \label{QVHE}

\subsection{The control algorithm realizing the QVHE acoustic analogue in real-time}  \label{Target_gen}

\begin{figure}[tbp]
\centering
\setlength{\tabcolsep}{1pt}
\begin{tabular}{c c c}
\setlength{\tabcolsep}{0.5pt}
\small \textbf{(a)} & \small \textbf{(b)}  & \small \textbf{(c)} \\
\includegraphics[height=3.2cm, valign=t]{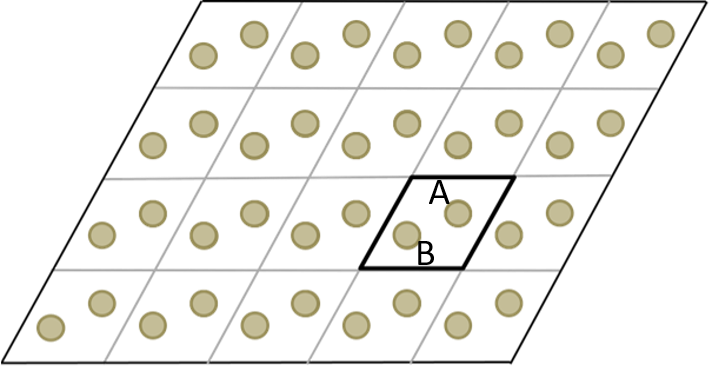} & \includegraphics[height=3.2cm, valign=t]{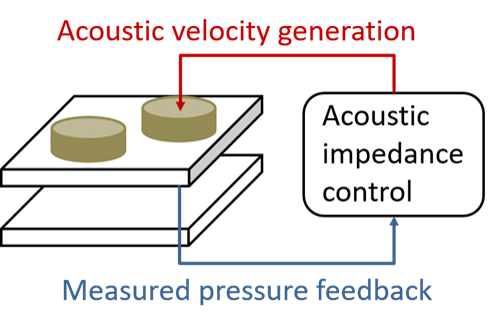} & \includegraphics[height=3.2cm, valign=t]{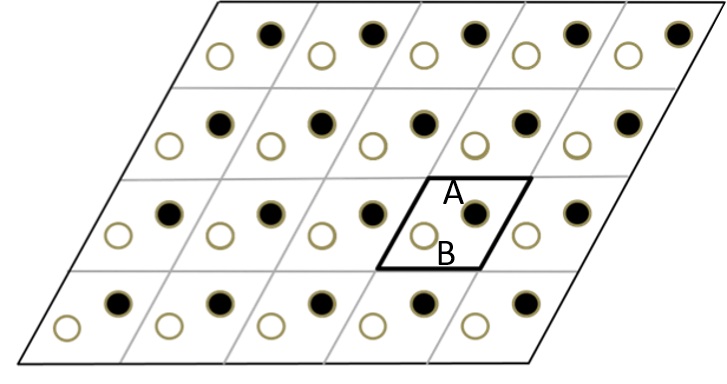} \\
\small \textbf{(d)} & \small \textbf{(e)}  & \small \textbf{(f)} \\
\includegraphics[height=3.2cm, valign=t]{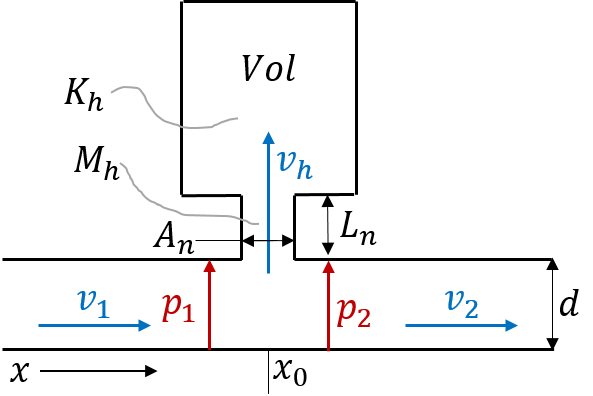} & \includegraphics[height=3.2cm, valign=t]{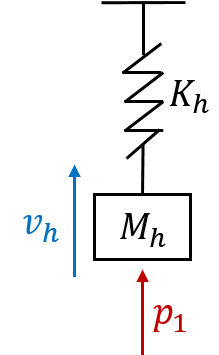} & \includegraphics[height=3.2cm, valign=t]{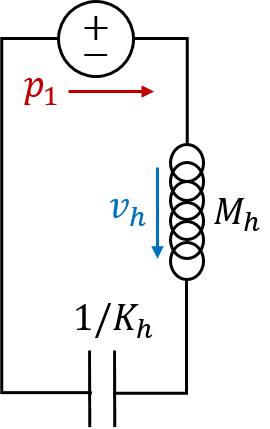}
\end{tabular}
\caption{Real-time closed-loop creation of the QVHE. (a) The mounting pattern of the all-identical acoustic actuators (gold circles). (b) The feedback control scheme at a single site. (c) A bipartite pattern of target acoustic impedances created by closed-loop control, \textit{imitating} acoustic resonators with respective impedances $Z_A$ (black circle) and $Z_B$ (open gold circle). (d) Schematic of the target acoustic (Helmholtz) resonator. (e) Mechanical and (f) electrical analogues of the acoustic resonator in (d).}
\label{Scheme}
\end{figure}

As was discussed in Sec. \ref{Intro}, the QVHE belongs to a family of effects, originating from quantum physics, that support topologically protected wave propagation. 
Here we outline the structural conditions for a classical analogue of this effect to take place. 
The most basic structure that is capable to support the QVHE is a uniform honeycomb lattice, as illustrated by the gold circles in Fig. \ref{Scheme}(a), for which a certain property is turned alternating between the sites, as captured by the gold and black circles in Fig. \ref{Scheme}(c). 
This transition involves breaking space inversion symmetry within a two-site unit cell, as outlined by the black parallelogram.
In quantum systems the lattice sites represent vacancies of alternating potential for hopping electrons \cite{franz2013topological}. The simplest classical mechanical analogy is a lattice of alternating masses connected by springs \cite{pal2017edge}. 
In a passive acoustic metamaterial the system in Fig. \ref{Scheme}(c) could be achieved e.g. by attaching Helmholtz resonators \cite{pierce1990acoustics,hu2019modelling} (cavities on necks, Fig. \ref{Scheme}(d)) of different geometries at sites $A$ and $B$ to one plate of a slab as in Fig. \ref{GenScheme}. This would result in alternating discrete acoustic impedances $Z$ at these locations, which is the ratio between the pressure field and the flow velocity at the resonator entrance \cite{pierce1990acoustics}. \\
In our system, however, no passive elements are included. Any changes of spatial symmetry are achieved only via active control of the pressure field. 
Figure \ref{Scheme}(a) therefore represents the upper plate of the waveguide in Fig. \ref{GenScheme}, with the uniform honeycomb pattern of the gold circles indicating the identical acoustic transducers.
We program these transducers to create, using real-time control, discrete changes of impedance, denoted by $Z_A$ and $Z_B$, at sites $A$ and $B$ of each unit cell in Fig. \ref{Scheme}(a), imitating Helmholtz resonators at these locations. 
We denote the locations of sites $A$ and $B$ by $\textbf{R}_A$ and $\textbf{R}_B$, respectively.
Since a Helmholtz resonator affects the acoustic impedance at its vicinity, imitating such a resonator by our control system implies controlling the acoustic impedance at each loudspeaker location.
This can be achieved by measuring the pressure with a microphone at that location, processing it through a controller, and generating the required flow velocity with the corresponding loudspeaker, as illustrated in Fig. \ref{Scheme}(b) for a single unit cell. Although the generated velocity is perpendicular to the propagation field in the waveguide, it is similarly coupled to the horizontal velocity field as with passive upper plate resonators. 
The control action in Fig. \ref{Scheme}(b) thereby converts the uniform pattern of Fig. \ref{Scheme}(a) to the target closed-loop metamaterial, which has the alternating pattern of Fig. \ref{Scheme}(c). 
The control law at each site $j$ is therefore a collocated pressure feedback, given by
\begin{equation}  \label{eq:Controller}
    v_j(s)=-H_j(s)p(\textbf{R}_j;s), \qquad H_j(s)=\frac{1}{Z_j(s)},
\end{equation}
with the controller $H_j(s)$ constituting the inverse of the desired local impedance $Z_j(s)$.
The $j$ unit cell of the closed-loop metamaterial, which we create using the feedback-based design, is then governed by
\begin{equation}  \label{eq:res_2D_2}
c^2\nabla^2p(\textbf{r};s)=s^2p(\textbf{r};s)+\rho c^2\eta \frac{s}{Z_A(s)}p(\textbf{R}_A;s)\delta(\textbf{r}-\textbf{R}_A)+\rho c^2\eta \frac{s}{Z_B(s)}p(\textbf{R}_B;s)\delta(\textbf{r}-\textbf{R}_B).
\end{equation}
This hybrid system, comprising a continuous two-dimensional pressure field augmented by discretely spanned changes of impedance, which is our target system in closed-loop, does not usually appear in acoustic textbooks in the form of \eqref{eq:res_2D_2}. It is mostly discussed for a one-dimensional waveguide, such as a tube, in the context of transmission and reflection of sound
through the resonator \cite{pierce1990acoustics}.
Here, however, the explicit form we derived in \eqref{eq:res_2D_2} is necessary to determine the parameters of the controllers in \eqref{eq:Controller}, since \eqref{eq:res_2D_2} can be analyzed for the dispersion properties, as carried out in Sec. \ref{QVHE}. \\
Since the impedances $Z_A(s)$ and $Z_B(s)$ are determined within the control program, their particular expressions can be arbitrary, up to hardware limits and causality. 
Here we use expressions corresponding to Helmholtz resonators.
As depicted in Fig. \ref{Scheme}(d), a Helmholtz resonator is a cavity of volume $Vol$ on a neck of length $L_n$ and area $A_n$, attached at location $x_0$ to the slab. 
This acoustic resonator is analogous to a mechanical mass-spring resonator (Fig. \ref{Scheme}(e)) of mass $M_h$ and spring constant $K_h$, with the impedance relating the force $p_1$ applied to the mass and its resulting velocity $v_h$. 
Another analogy, with which the notion of impedance is naturally associated, is an electrical $LC$ circuit (Fig. \ref{Scheme}(f)), of inductance $M_h$ and capacitance $1/K_h$, with the impedance relating the voltage $p_1$ and the current $v_h$. 
In the acoustic resonator the impedance relates the pressure $p_1$ and the flow velocity $v_h$ at the resonator entrance. For each $A$ and $B$ site, the impedance of our target closed-loop acoustic metamaterial is given by
\begin{equation}  \label{eq:Z_12}
    Z_A(s)=M_As+D_h+\frac{K_A}{s}, \qquad Z_B(s)=M_Bs+D_h+\frac{K_B}{s}. 
\end{equation}
Here $D_h$ represents dissipation, whereas $M_{A,B}=M_h(1+\epsilon_M)$ and $K_{A,B}=K_h(1+ \epsilon_K)$ are the equivalent acoustic mass and spring constant, respectively determined by the neck and cavity parameters, as $M_h=\rho L_n$ $[kg/m^2]$ and $K_h=A_n\rho c^2/Vol$ $[N/m^3]$. $\epsilon_M\in (-1,1)$ and $\epsilon_K\in (-1,1)$ are free design parameters indicating possible deviations from the nominal value, which can be used to achieve the desired pattern alternation in Fig. \ref{Scheme}(c).
The controllers in \eqref{eq:Controller} that realize the metamaterial in \eqref{eq:res_2D_2} in real-time with the target impedance relations in \eqref{eq:Z_12} thus take the leading phase form
\begin{equation}  \label{eq:Controller_H}
    H_A(s)=\frac{s}{M_As^2+D_hs+K_A} \qquad , \qquad H_B(s)=\frac{s}{M_Bs^2+D_hs+K_B}.
\end{equation}
The resulting closed-loop system is stable as long as the target resonators include damping, i.e. $D_h>0$. 
Next we calculate the dispersion relation of \eqref{eq:res_2D_2}-\eqref{eq:Z_12} to analyze its topological properties, and to determine the conditions required for the QVHE acoustic analogue to be supported in this system.

\subsection{Dispersion analysis of the closed-loop metamaterial that imitates the QVHE}   \label{Target}

\subsubsection{Dispersion characteristics of the bulk metamaterial}   \label{Inf_an}

\begin{figure}[bp]
\centering
\begin{tabular}{c c c}
\small \textbf{(a)} & \small \textbf{(b)}  & \small \textbf{(c)} \\
\includegraphics[width=5.4cm, valign=c]{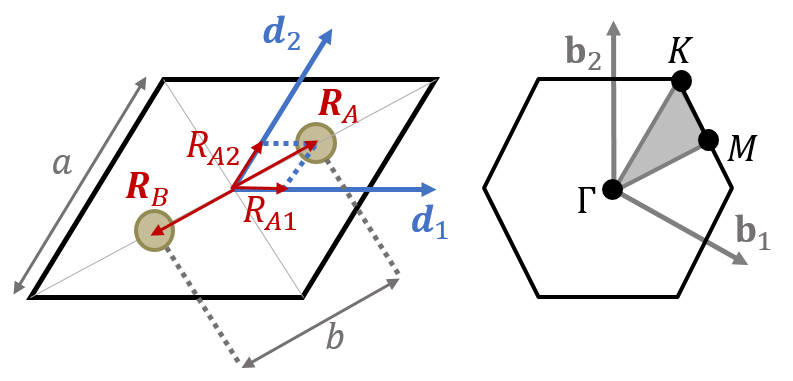} & \includegraphics[width=5.4cm, valign=c]{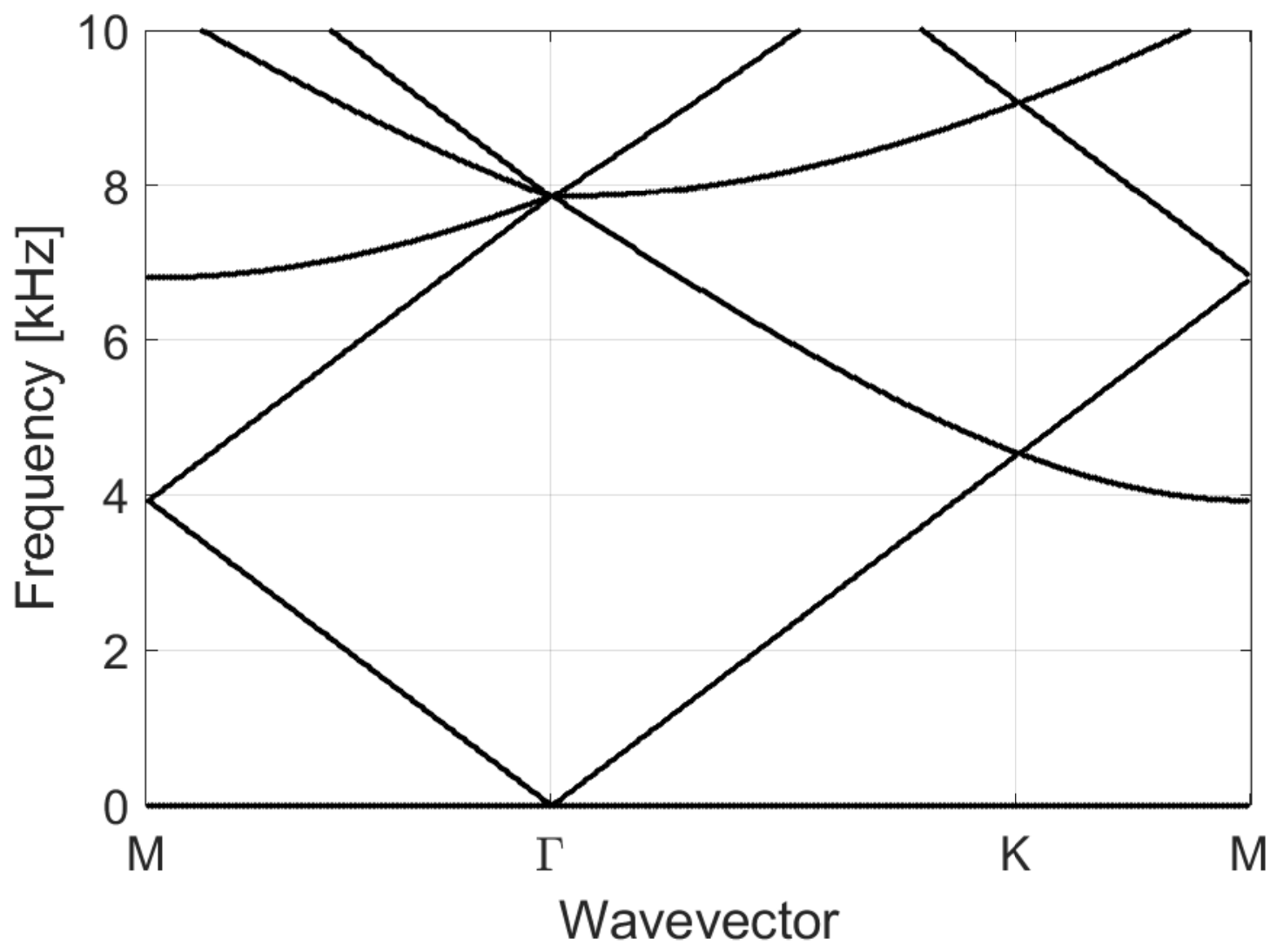} & \includegraphics[width=5.4cm, valign=c]{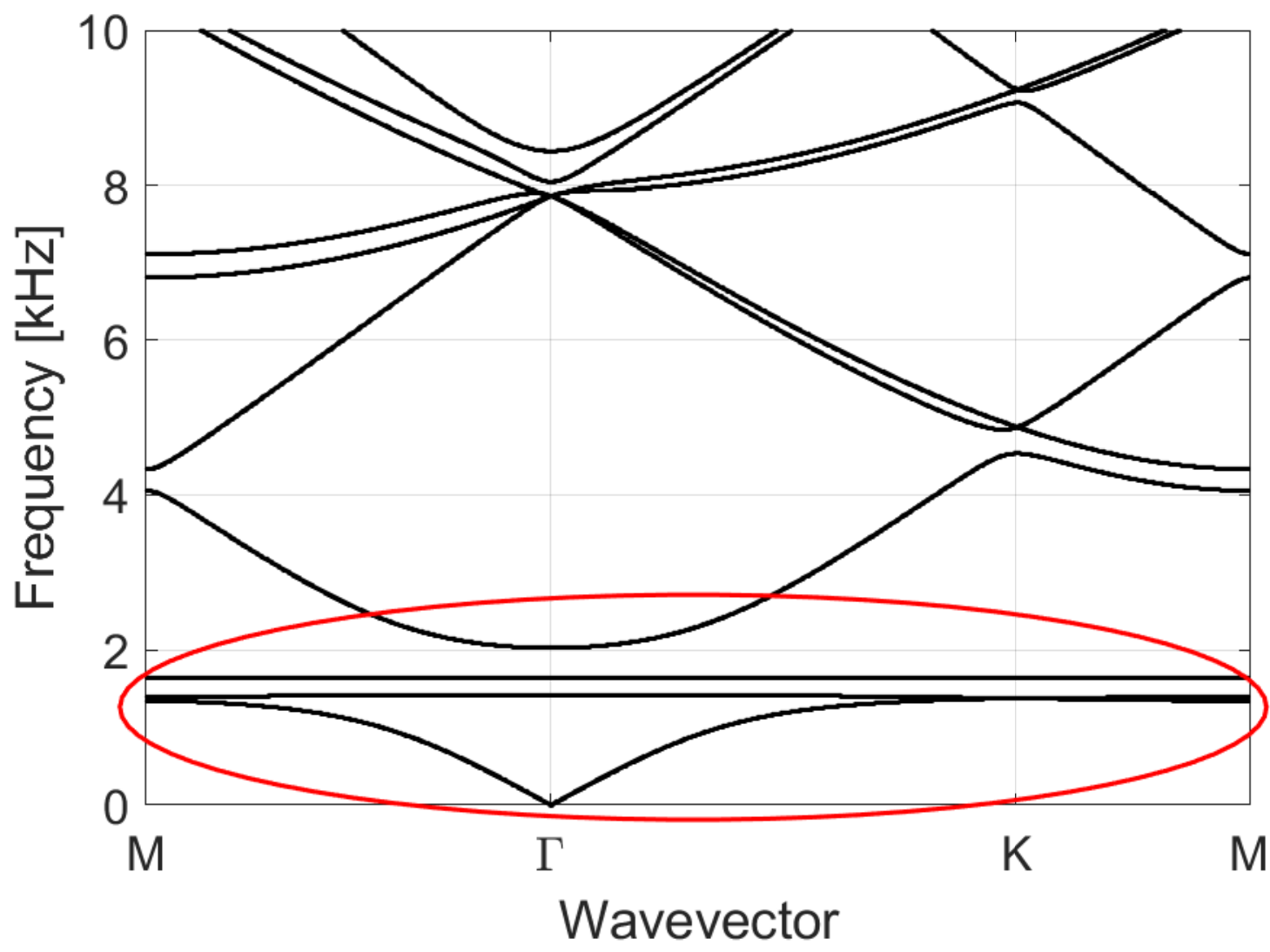} \\
\small \textbf{(d)} & \small \textbf{(e)}  & \small \textbf{(f)}  \\
\includegraphics[width=5.4cm, valign=c]{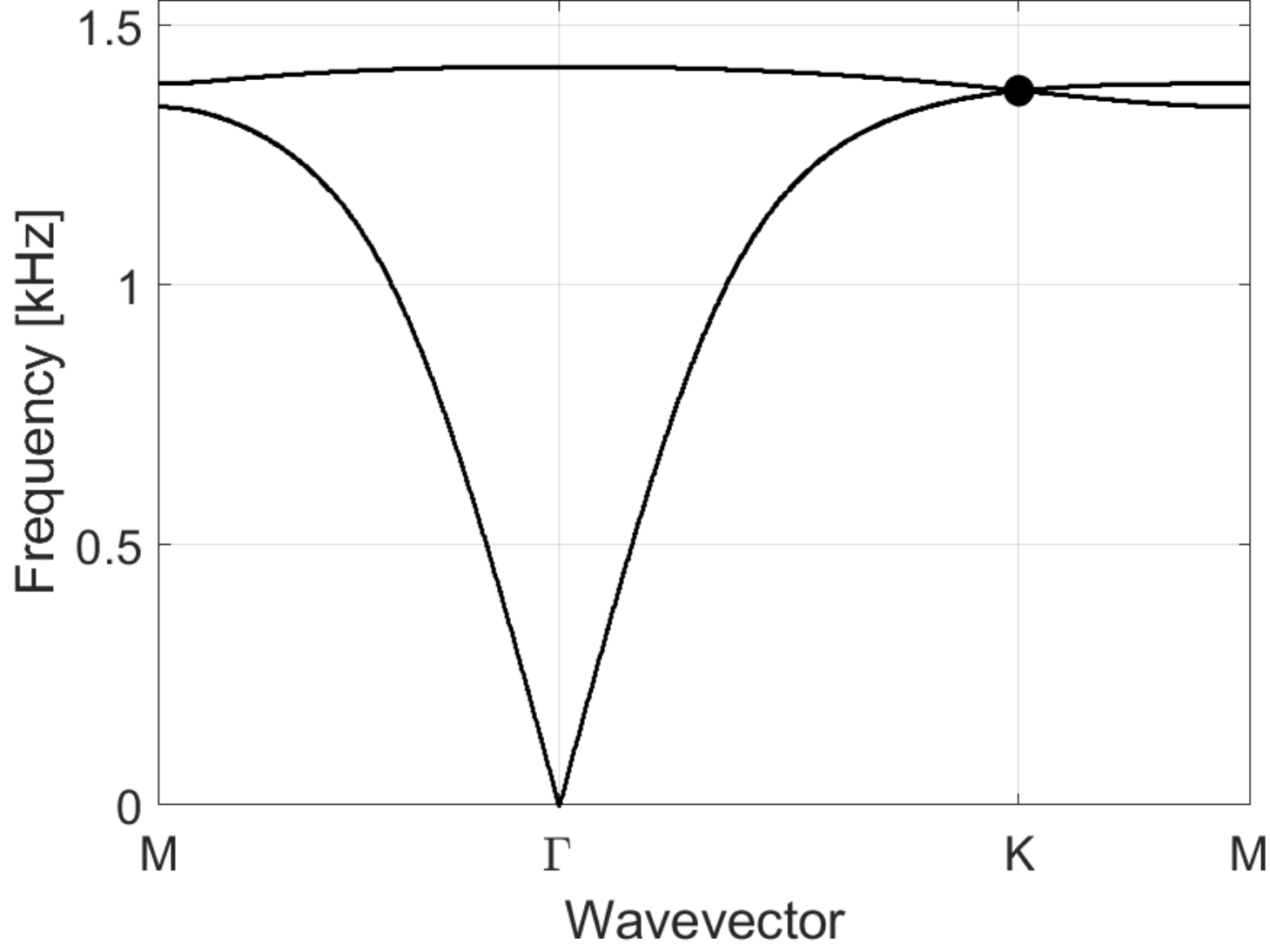} & \includegraphics[width=5.4cm, valign=c]{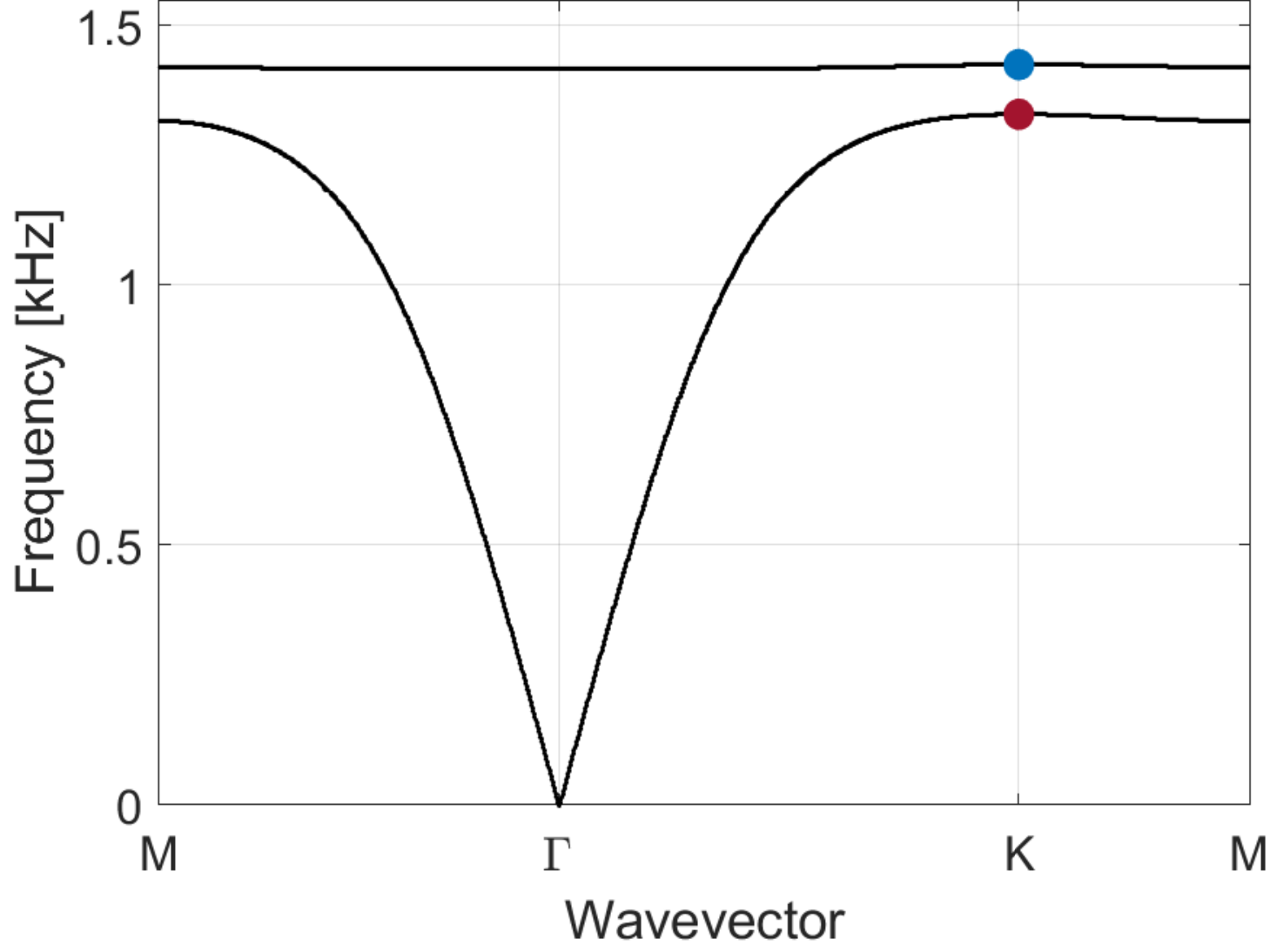} & \includegraphics[width=5.4cm, valign=c]{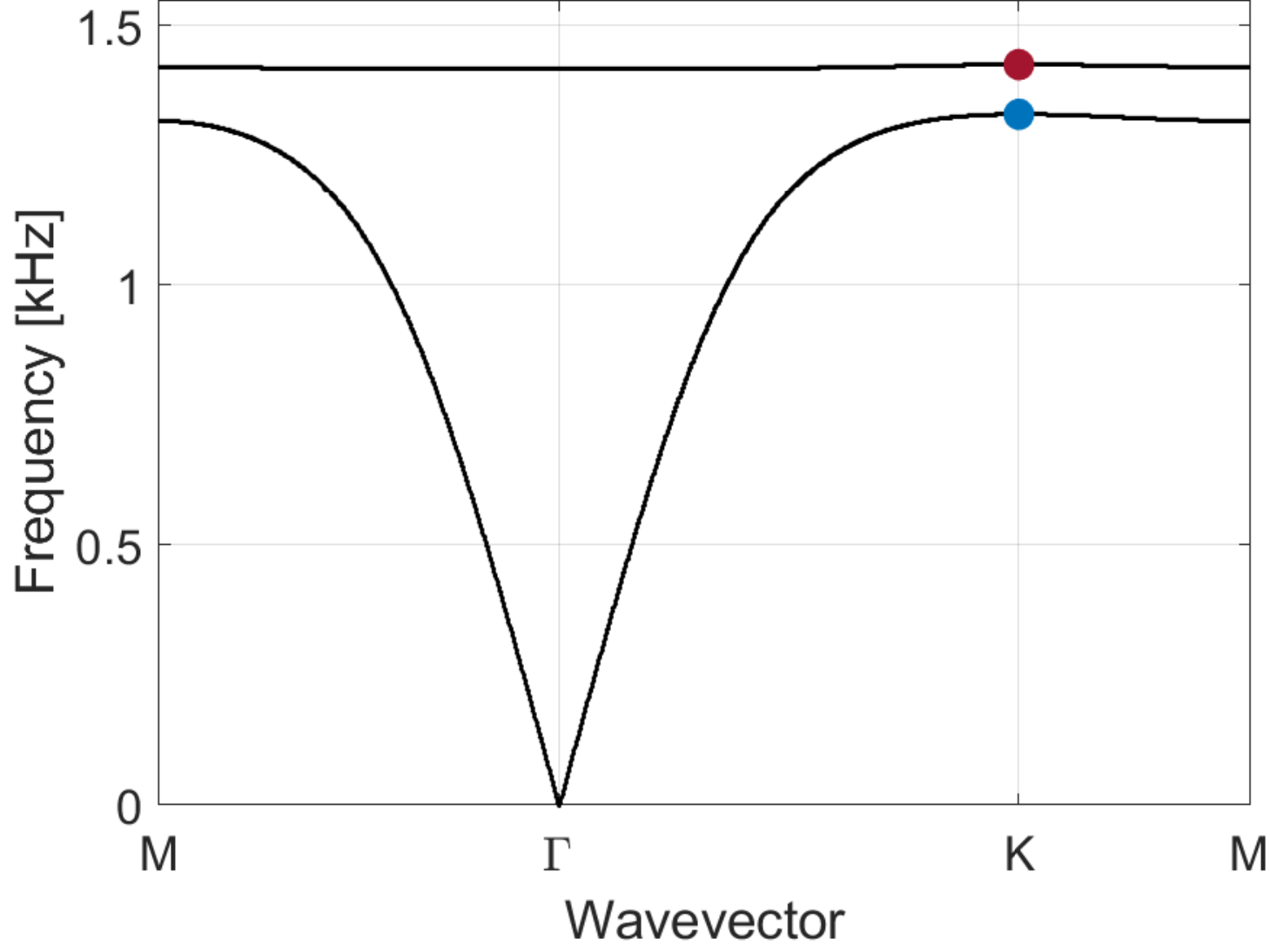} \\
\small \textbf{(g)} & \small \textbf{(h)}  & \small \textbf{(i)} \\
\includegraphics[height=1.2cm, valign=c]{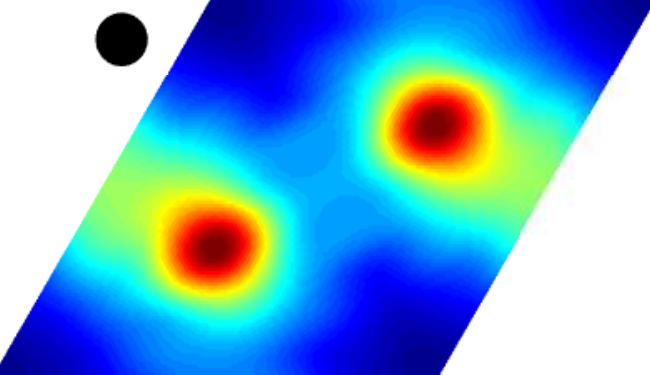} & \includegraphics[height=1.2cm, valign=c]{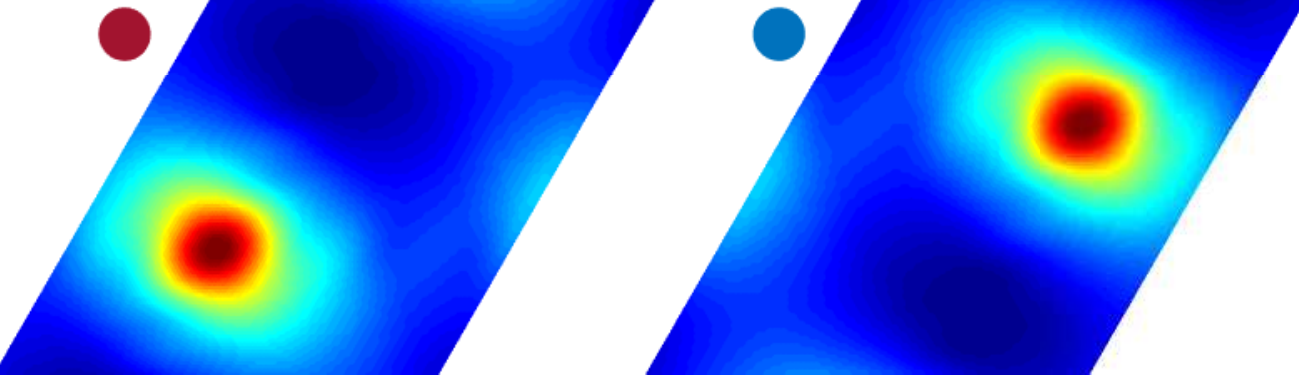}  & \includegraphics[height=1.2cm, valign=c]{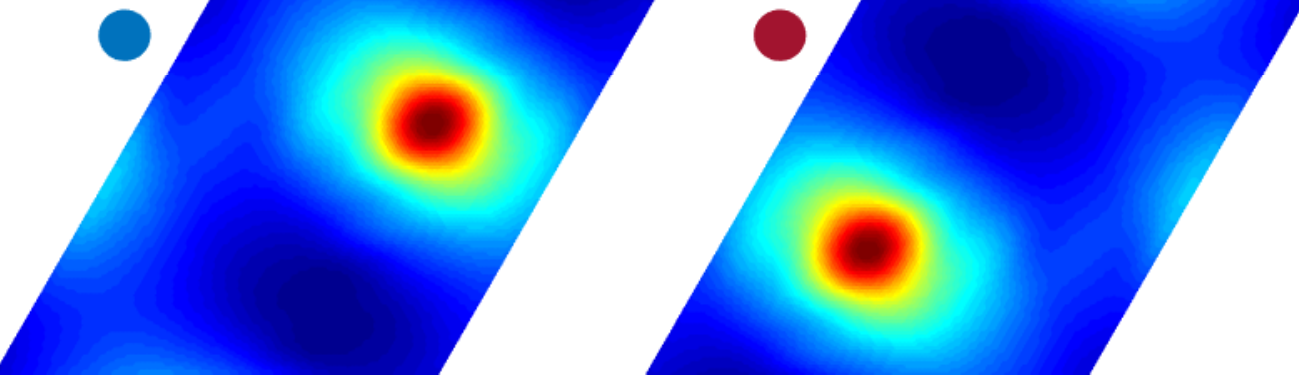}
\end{tabular}
\caption{Dispersion relation of an infinite hybrid continuous-discrete closed-loop metamaterial. (a) Left - unit cell. Right - the first Brillouin zone. (b) Dispersion relation of the metamaterial when control is off (the slab waveguide). (c) Dispersion relation and (g) eigenmode at the $K$ point of the metamaterial when control is turned on and creates a honeycomb pattern of identical impedances. (d) Zoom-in at the low frequency bands in (c), which are decoupled from the higher bands due to control. (e) Low frequency dispersion and (h) eigenmodes at the $K$ point with $M_A=1.1M_h$, $M_B=0.9M_h$. A gap is opened between the bands. (f) Low frequency dispersion and (i) eigenmodes at the $K$ point with $M_A=0.9M_h$, $M_B=1.1M_h$. The dispersion profile is identical to (e), but the modes are flipped, indicating a topological transition.}
\label{Bandstructure_infinite}
\end{figure}

Since the model in \eqref{eq:res_2D_2} is a hybridization of a continuous acoustic pressure field and discretely spaced impedance changes, the frequency dispersion with wavevector, $\omega(\textbf{k})$, cannot be calculated directly using a traveling harmonic wave solution $e^{i(\textbf{k}\cdot\textbf{r}-\omega t)}$.
We therefore invoke a semi-analytical approach, denoted by the Plane Wave Expansion method \cite{chaunsali2018subwavelength,xiao2012flexural,wang2015tuning}, 
to calculate the frequency dispersion of the infinite periodic metamaterial of Figs. \ref{Scheme}(a,c). 
The unit cell is depicted in Fig. \ref{Bandstructure_infinite}(a)-left, where the gold circles indicate the locations of the target resonators, given by $\textbf{R}_A=R_{A1}\textbf{d}_1+R_{A2}\textbf{d}_2$ and $\textbf{R}_B=-\textbf{R}_A$ in the real lattice space.
The wavevector $\textbf{k}=k_1\textbf{b}_1+k_2\textbf{b}_2$ is evaluated in the reciprocal lattice space along a one-dimensional path connecting the high symmetry points $M - \Gamma - K - M$, as illustrated in Fig. \ref{Bandstructure_infinite}(a)-right, enclosing the Irreducible Brillouin Zone (the shaded area) \cite{brillouin1953wave}. 
The calculation details of frequency evolution with respect to the wavevector as the solutions of an augmented eigenvalue problem appear in Appendix \ref{PWE}. \\
First, we consider the metamaterial in Fig. \ref{Scheme}(a) for the trivial open-loop (uncontrolled) case, which indicates a bare waveguide without any changes of impedance, obtained for $Z_A(s)=Z_B(s)\rightarrow\infty$ in \eqref{eq:res_2D_2}.
This is a fully continuous system with a standard dispersion of the two-dimensional wave equation, which is indeed retrieved by our calculation, as appears in Fig. \ref{Bandstructure_infinite}(b).
As expected for a uniform impedance system, the dispersion curve, folded here into the Irreducible Brillouin Zone, is gapless (truncated at 10 $[kHz]$ in the figure), i.e. traveling waves are supported at any temporal frequency. \\
Next we consider a waveguide with identical impedance changes $Z_A(s)=Z_B(s)$ in every unit cell of the honeycomb pattern in Fig. \ref{Scheme}(c) with $a=0.05$ $[m]$, imitating identical Helmholtz resonators (defined in \eqref{eq:Z_12}) for $Vol=4.917\cdot 10^{-6}$ $[m^3]$, $A_n=4.524\cdot 10^{-6}$ $[m^2]$ and $L_n=10^{-3}$ $[m]$. The distance between plates in the calculation of $\eta$ in \eqref{eq:res_2D_2} was $d=5\cdot 10^{-3}$ $[m]$.
The system is now a continuous-discrete hybridization.
Remarkably, introducing the periodically spanned change of impedance, even when space inversion symmetry within the unit cell is preserved, decouples the lower frequency bands from the higher bands, as illustrated in Fig. \ref{Bandstructure_infinite}(c). 
This allows for an isolated working regime, which is encircled in red in Fig. \ref{Bandstructure_infinite}(c) and enlarged in Fig. \ref{Bandstructure_infinite}(d). 
The low frequency regime consists of two bands connected at a single (Dirac) point $K$, marked by a black dot in the figure. 
This two-band wave dispersion diagram resembles the dispersion of a purely discrete bipartite lattice with a single degree of freedom per site, for which the two bands constitute the entire spectrum, such as in the electronic band-structure of graphene \cite{reich2002tight}. 
The corresponding eigenmode at the $K$ point is depicted in Fig. \ref{Bandstructure_infinite}(g), indicating the pressure field distribution in the unit cell. Since the target resonators in this case are identical, the pressure distribution is equal at both $\textbf{R}_A$ and $\textbf{R}_B$. \\
Our ultimate goal, however, is target resonators of different impedances $Z_A(s)\neq Z_B(s)$, implying space inversion symmetry breaking in the unit cell, which we achieve here by the embedded feedback control operation.
For Helmholtz resonators, the impedance of which is of the form given in \eqref{eq:Z_12}, the difference can be obtained either by imitating $K_A\neq K_B$, or $M_A\neq M_B$, or both. Assuming, for example, $M_A\neq M_B$ and $K_A=K_B$, we set $\epsilon_K=0$ and $\epsilon_M=\epsilon\in (-1,1)$ in the definitions below \eqref{eq:Z_12}. 
We performed the calculation for two values, $\epsilon=0.1$ and $\epsilon=-0.1$. The resulting dispersion relations are shown in Fig. \ref{Bandstructure_infinite}(e) and (f), respectively. 
In both cases a gap is opened between the two bands at the $K$ point, indicating that for frequencies within the gap, wave propagation is not supported in the metamaterial bulk. \\
The appearance of a gap due to space inversion symmetry breaking is a central feature in topological systems, as discussed next.
Since the transition between the gapped states occurs only through a gap closing, together with a nonzero topological invariant carried by each band (see Appendix \ref{Top_inv}), the systems corresponding to $\epsilon>0$ and to $\epsilon<0$ are topologically different.
Although the gapped dispersion profile is identical for both values of $\epsilon$ (since for an infinite system flipping the masses is essentially equivalent to translation), the transition between the two systems is still captured in the dispersion data. Specifically, it is captured by the eigenmodes corresponding to the two frequencies at the $K$ point, which are labeled by the red and blue dots in Figs. \ref{Bandstructure_infinite}(e) and (f). When the value of $\epsilon$ is flipped, the eigenmodes, which are respectively depicted in Fig. \ref{Bandstructure_infinite}(h) and (i), are flipped as well. 
Topological wave propagation, which is the goal of our metamaterial design, is obtained on an interface of the topologically different $\epsilon>0$ and $\epsilon<0$ systems.
To obtain the interface states in the dispersion relation, we consider a periodic metamaterial with an extended unit cell, which contains the interface, as outlined in Sec. \ref{Edge_disp}.

\subsubsection{Dispersion characteristics of a finite-sized metamaterial}   \label{Edge_disp}

\begin{figure}[htpb]
\centering
\begin{tabular}{l l l l l l l}
\textbf{(a)} & \includegraphics[height=4.9cm, valign=t]{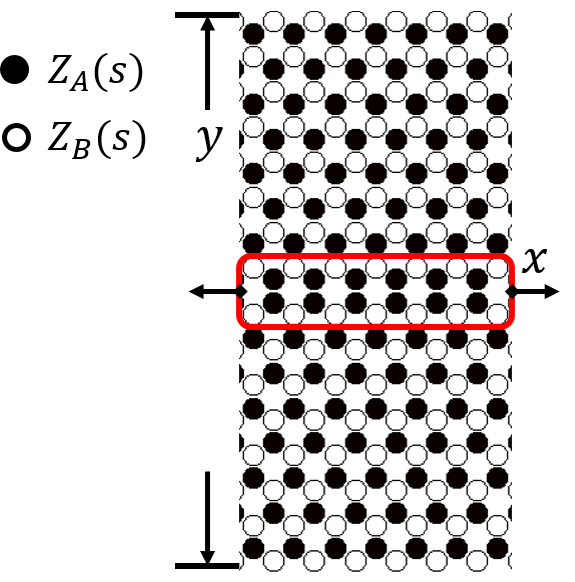} & \textbf{(b)} & \includegraphics[height=5.3cm, valign=t]{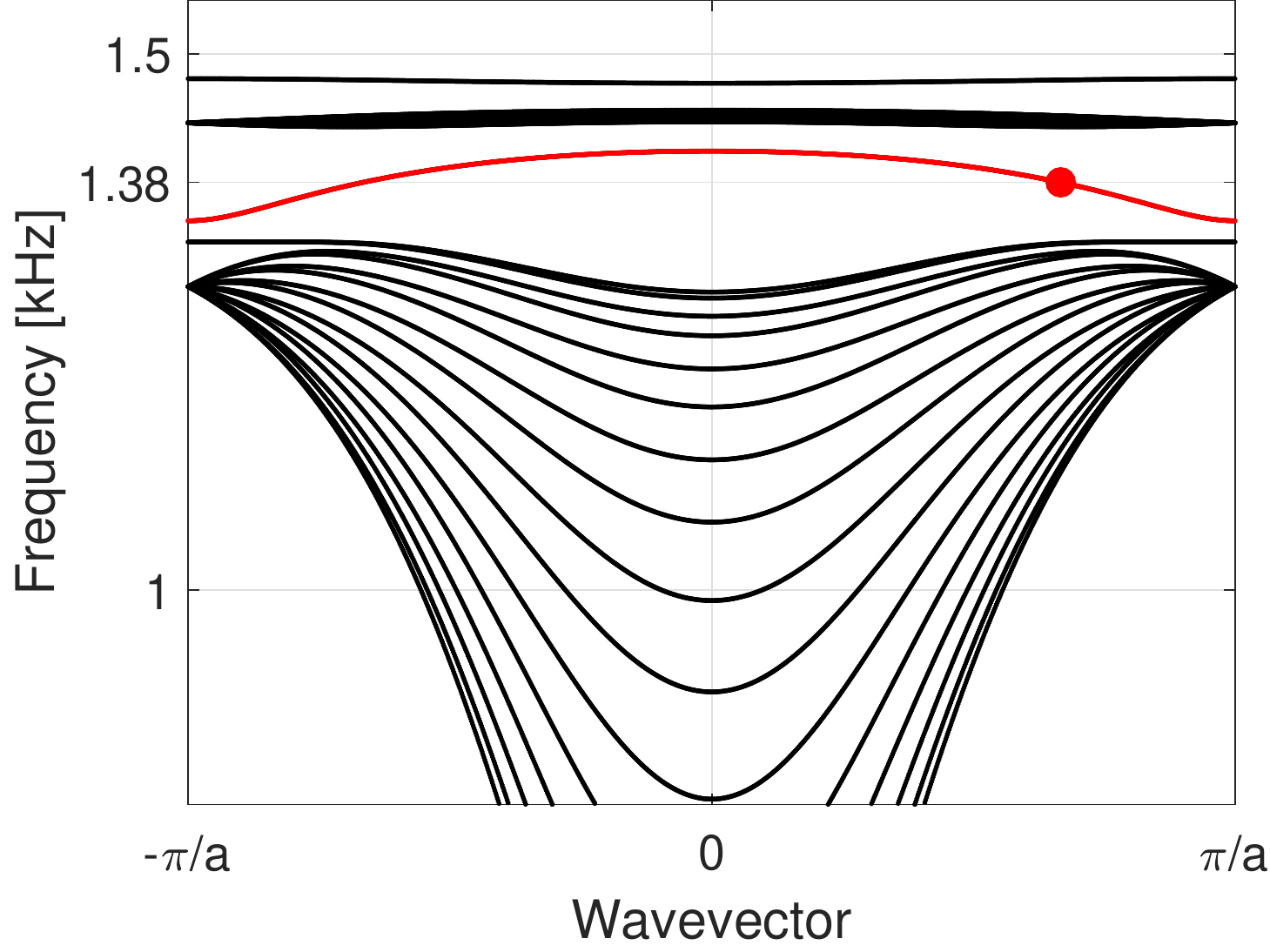} & \textbf{(c)} \includegraphics[height=4.8cm, valign=t]{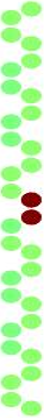} &  \includegraphics[height=2.5cm, valign=t]{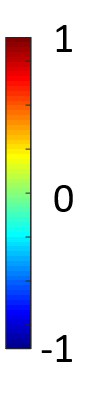}
\end{tabular}
\caption{Band-structure of an equivalent discrete target metamaterial, infinite-periodic in the $x$ direction and finite in the $y$ direction. (a) Lattice schematic, comprising $Z_A$ (black circles) and $Z_B$ (white circles) impedances, with an interface (red). (b) The band-structure, calculated for a 16 two-site cells strip, which includes bulk states (black) and an interface state (red). (c) The corresponding eigenmodes at the red dot.}
\label{Bandstructure_edge}
\end{figure}

{As was discussed in Sec. \ref{Inf_an} and captured by Fig. \ref{Bandstructure_infinite}(e) and (f), when the impedance at sites $A$ and $B$ is different, a gap opens between the two decoupled low frequency dispersion states of the infinite periodic system. }
As a result, harmonic waves of frequencies that lie within the gap cannot propagate in the bulk, implying that at these frequencies the system behaves as an insulator for acoustic waves. %
In this section we investigate the effect of an interface of identical impedance at sites $A$ and $B$ on the frequency dispersion. We consider a metamaterial that is infinite along the $x$ axis and finite along the $y$ axis, spanned by a periodic pattern of target closed-loop resonators, as illustrated in Fig. \ref{Bandstructure_edge}(a). 
This pattern is periodic in the $x$ axis with each vertical strip constituting an extended unit cell, or a super-cell.
The black and white circles respectively indicate impedances $Z_A(s)$ and $Z_B(s)$, as defined in \eqref{eq:Z_12}, created in real-time by the embedded control system. 
Each super-cell contains an interface of identical impedance, here, for example, $Z_A(s)$, encircled in red in Fig. \ref{Bandstructure_edge}(a). \\
Calculating the dispersion relation of this system by the Plane Wave Expansion method, as we did for the fully infinite system in Appendix \ref{PWE}, now becomes more involved, as we need to distinguish between spatial derivatives $p_{xx}$ and $p_{yy}$ in \eqref{eq:res_2D_2}. Since the $y$ coordinate is now finite, $p_{yy}$ will not be eliminated,
but will yield a polynomial eigenvalue problem, which is also differential. 
The common resort in calculating dispersion of semi-infinite systems is the Finite Element method \cite{chaunsali2018subwavelength}.
We take a different approach by developing an equivalent tractable model of a purely discrete system, the sites of which coincide with the pattern in Fig. \ref{Bandstructure_edge}(a), thus preserving the insight of the analytical treatment.
As detailed in Appendix \ref{Open}, we adjust the parameters of the equivalent system until we obtain an exceptional fitting of the dispersion relation with the original hybrid continuous-discrete system in the fully infinite configuration. 
We therefore regard all calculations performed on the equivalent model as absolute representatives of the original model.
The resulting frequency dispersion for the super-cell of thirty two sites (comprising sixteen primitive two-site cells) is presented in Fig. \ref{Bandstructure_edge}(b). 
The bulk states are plotted in black. \\
We observe that a band gap is formed exactly at the same frequency region as for the fully infinite configuration (Fig. \ref{Bandstructure_infinite}(e) and (f)). In addition, a new state, which is plotted in red, emerges inside the gap. This state corresponds to the interface of identical impedances.
The dispersion plot for a $Z_A(s)$ interface case is quite similar to Fig. \ref{Bandstructure_edge}(b), yet not identical, as such system cannot be converted into the one with a $Z_B(s)$ interface by a simple translation. 
Since the dispersion of the bulk metamaterial is gapped at the region of the interface state, waves of frequencies at that region can propagate only along the interface.
Due to the topological property of the dispersion relation (evaluated for the infinite configuration in Appendix \ref{Top_inv}), these waves are expected to be strictly confined to the interface, and to remain localized on it in the presence of sharp turns and corners.
This expectation is validated by the corresponding eigenmode (calculated for $\omega=1.38$ $[kHz]$), depicted in Fig. \ref{Bandstructure_edge}(c), which is clearly localized on the interface. 
The resulting time domain wave propagation is demonstrated in Sec. \ref{Time_sim}.

\subsection{{A control algorithm realizing a different topological effect in real-time}}  \label{Target_QHE}

{Here we present, without an accompanying dispersion analysis, a control algorithm that realizes an acoustic analogue of a different quantum topological phenomenon on the same platform of Fig. \ref{GenScheme}. This is to demonstrate that our feedback-based design is not limited to a single effect, such as the QVHE, obtained by a relatively simple collocated pressure feedback control \eqref{eq:Controller}, but rather can realize other effects involving more complicated, not necessarily collocated control laws. 
For example, we would like our autonomous acoustic metamaterial not only to support guiding of curved sound beams, as enabled by the algorithm in Sec. \ref{Target_gen}, but also to make these beams unidirectional, which is even more unconventional for propagation of sound in free two-dimensional space. 
To this end, we design the following controller
\begin{equation}  \label{eq:Cont_law_QHE}
   \left[\begin{array}{c} v^A_{q,l}(s) \\ v^B_{q,l}(s) \end{array}\right]=-H_{q,l}(s)\left[\begin{array}{c} p^A_{q+ 1,l}+p^A_{q- 1,l+ 1}+p^A_{q,l- 1} +  p^A_{q- 1,l}+p^A_{q+ 1,l- 1}+p^A_{q,l+ 1} \\ p^B_{q+ 1,l}+p^B_{q- 1,l+ 1}+p^B_{q,l- 1} +  p^B_{q- 1,l}+p^B_{q+ 1,l- 1}+p^B_{q,l+ 1} \\ p^A_{q+ 1,l}+p^A_{q- 1,l+ 1}+p^A_{q,l- 1} -  p^A_{q- 1,l}-p^A_{q+ 1,l- 1}-p^A_{q,l+ 1} \\ p^B_{q+ 1,l}+p^B_{q- 1,l+ 1}+p^B_{q,l- 1} -  p^B_{q- 1,l}-p^B_{q+ 1,l- 1}-p^B_{q,l+ 1} \end{array}\right], \quad H_{q,l}(s)=\frac{1}{B_0\eta}\left[\begin{array}{c c c c}
\frac{1}{s}h_1 & 0 & h_2 & 0 \\
0 & \frac{1}{s}h_1 & 0 & -h_2 
\end{array}\right],
\end{equation}
which relates the acoustic velocity actuators of each site in Fig. \ref{Scheme}(a) to pressure measurements at adjacent sites (the indices $q$ and $l$ correspond to the principal coordinates $\textbf{d}_1$ and $\textbf{d}_2$, respectively).
We set the feedback gains as $h_1=\beta B_0/M_0 \cos \phi$ and $h_2=\beta B_0/M_0 \sin \phi$, where $\beta\in (0,1)$ and $\phi\in (0,\pi)$. $B_0=\rho c^2/b$ and $M_0=\rho b$ indicate the effective bulk modulus and mass density of air between neighboring sites, where $b=a/\sqrt{3}$ is the distance between them. 
It is important to note that although the acoustic velocity inputs $v_j(s)$ in the open loop (uncontrolled) system enter the equation \eqref{eq:res_2D_OL} through a differentiating operator $s$, the transfer functions from these inputs to the pressure field $p(\textbf{r};s)$ do not contain a zero at the origin. This is analytically proved for one-dimensional acoustic waveguides \cite{sirota2019active,sirota2020modeling}. In fact, the transfer function from velocity to pressure in an acoustic waveguide is similar to the transfer function from force to velocity in a mechanical flexible structure that is governed by the second order wave equation \cite{sirota2010free,sirota2015fractional,sirota2015fractionalA}. The integrator component in the controller in \eqref{eq:Cont_law_QHE} therefore does not cause any unstable pole-zero cancellations. We verify the closed-loop system stability for the set of parameters that is of interest. 
The control law in \eqref{eq:Cont_law_QHE} creates new direct couplings between all $A$ sites, and independently, between all $B$ sites.
The velocity flow along these new artificial channels is directional, which breaks time reversal symmetry of the system \cite{haldane1988model,wang2015topological,nash2015topological}. The resulting closed-loop metamaterial realizes an acoustic analogue of the quantum Hall effect \cite{haldane1988model}. For the controller gain ranges indicated above, the dispersion profile of the closed-loop metamaterial is characterized by a nontrivial topological invariant (specifically, $+1$ or $-1$) \cite{franz2013topological}. The sign of the phase parameter $\phi$ indicates the edge wave propagation direction, with positive (negative) for counterclockwise (clockwise). A prototype feedback-based realization in a purely mechanical mass-spring lattice with a detailed dispersion analysis appears in \cite{sirota2020feedback}. 
In time domain, a strictly uni-directional, topologically protected propagation of acoustic beams is then supported along the outer edges of the slab in Fig. \ref{GenScheme}, and between any domains with different topological invariants, as we demonstrate in Sec. \ref{Time_sim}.}

\section{Dynamical simulations of the metamaterial demonstrating topologically protected wave propagation}    \label{Time_sim}

\begin{figure}[htpb]
\centering
\setlength{\tabcolsep}{0pt}
\begin{tabular}{c | c | c}
\textbf{(a)} Control program 1 & \textbf{(e)} Control program 2 & \textbf{(i)} Control program 3 \\
\includegraphics[height=4.3cm, valign=c]{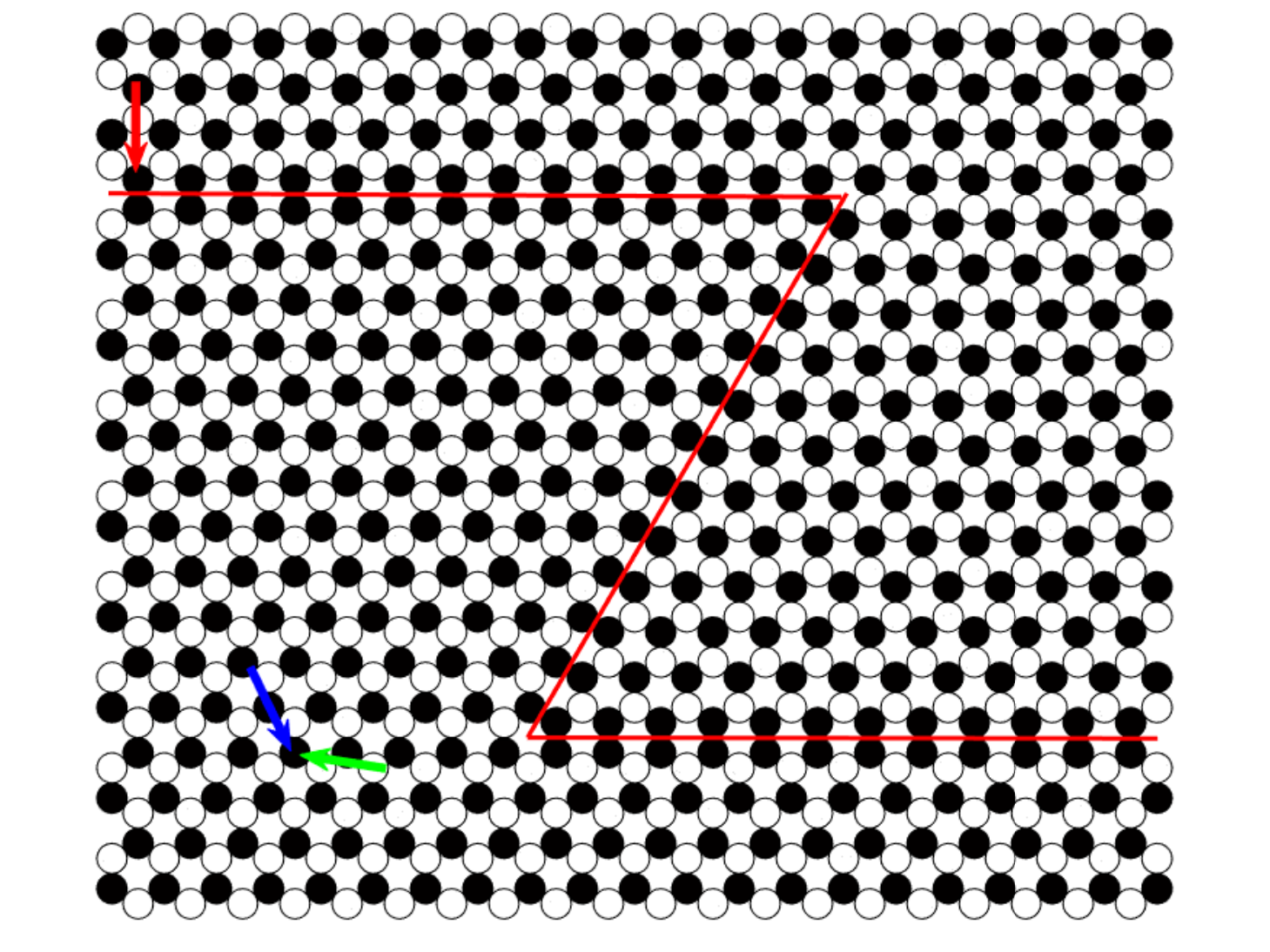}  & \includegraphics[height=4.3cm, valign=c]{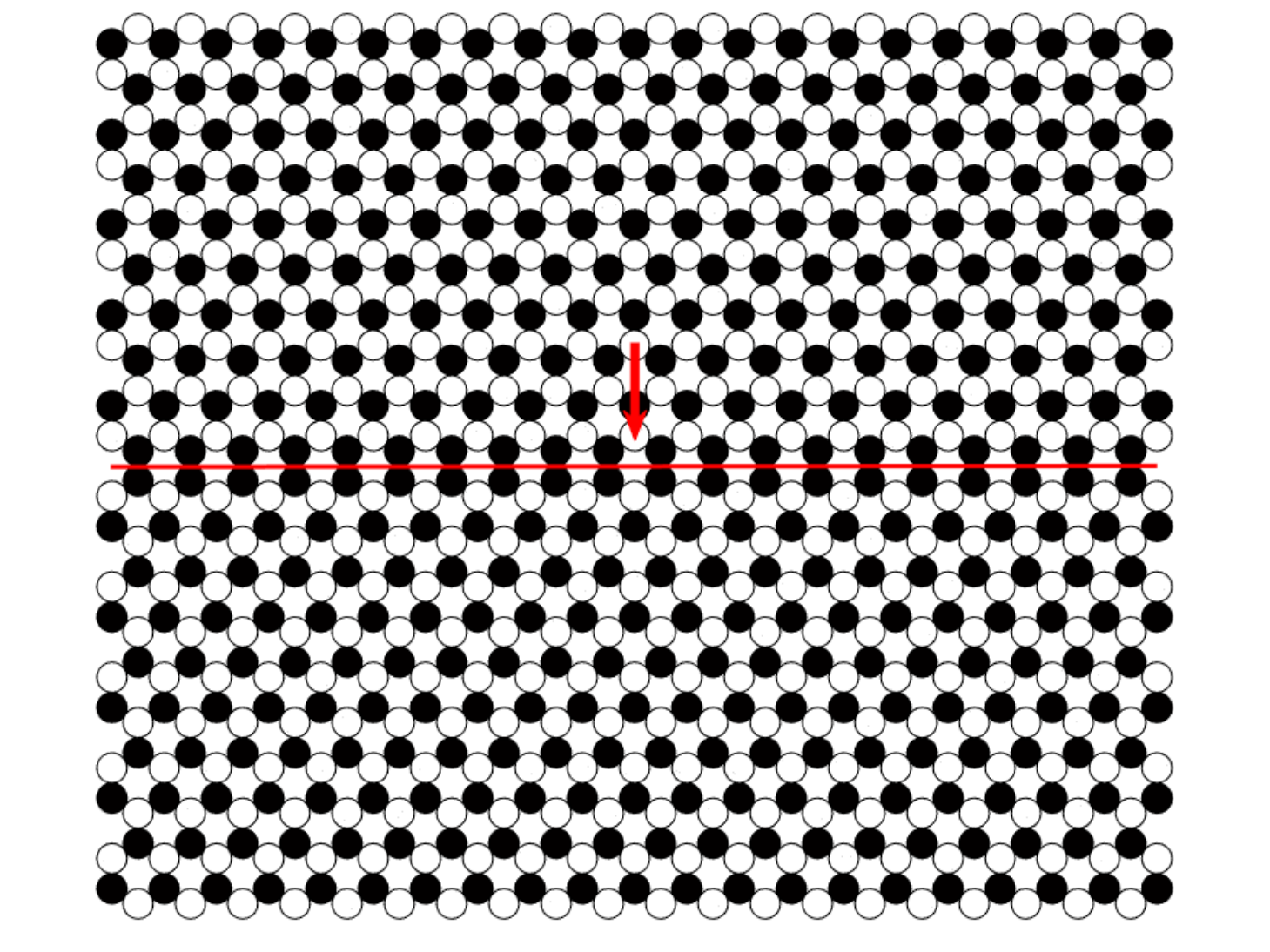} & \includegraphics[height=4.3cm, valign=c]{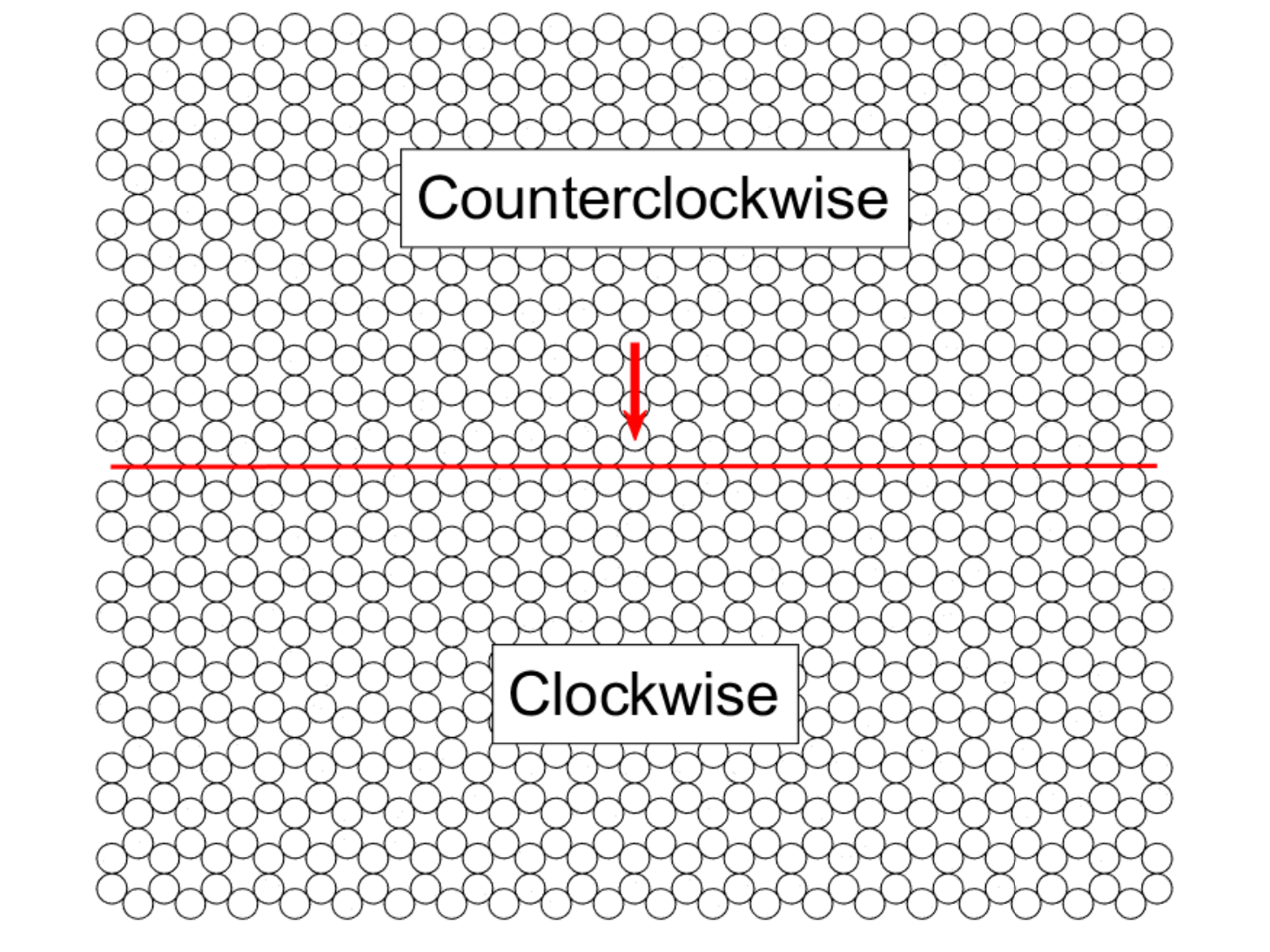} \\
\textbf{(b)} $f_1=1.38$ ${[kHz]}$ (red arrow) & \textbf{(f)} $f_4=1.38$ ${[kHz]}$, $T_1$ & \textbf{(j)} $f_5=2.89$ ${[kHz]}$, $T_1$ \\
\includegraphics[height=4.3cm, valign=c]{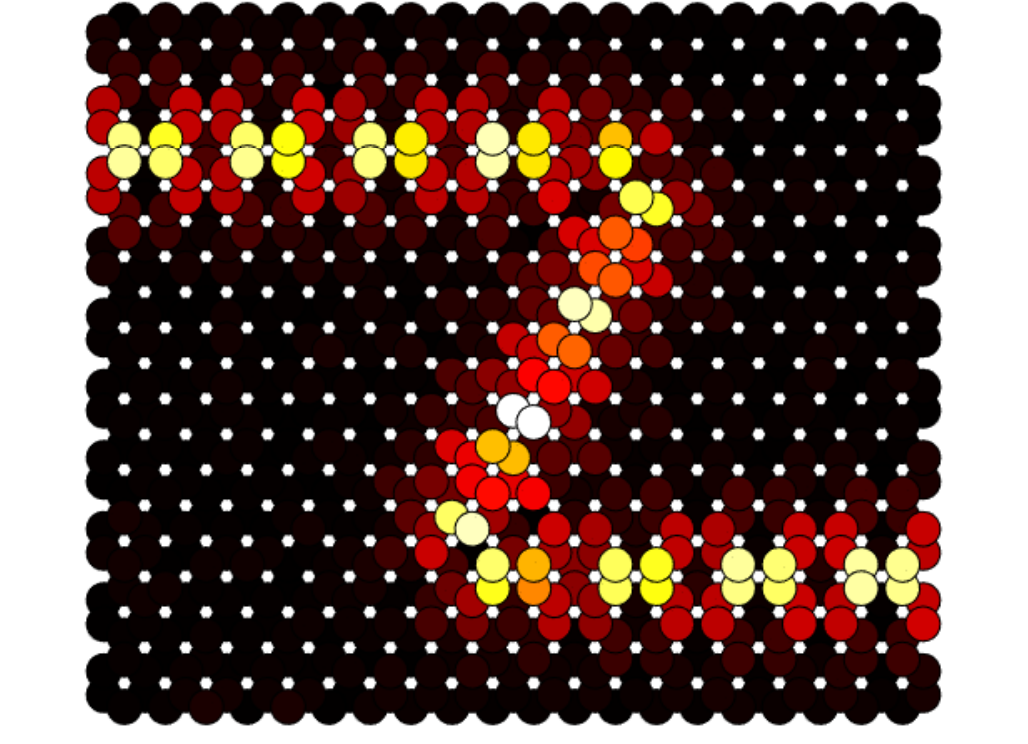} & \includegraphics[height=4.3cm, valign=c]{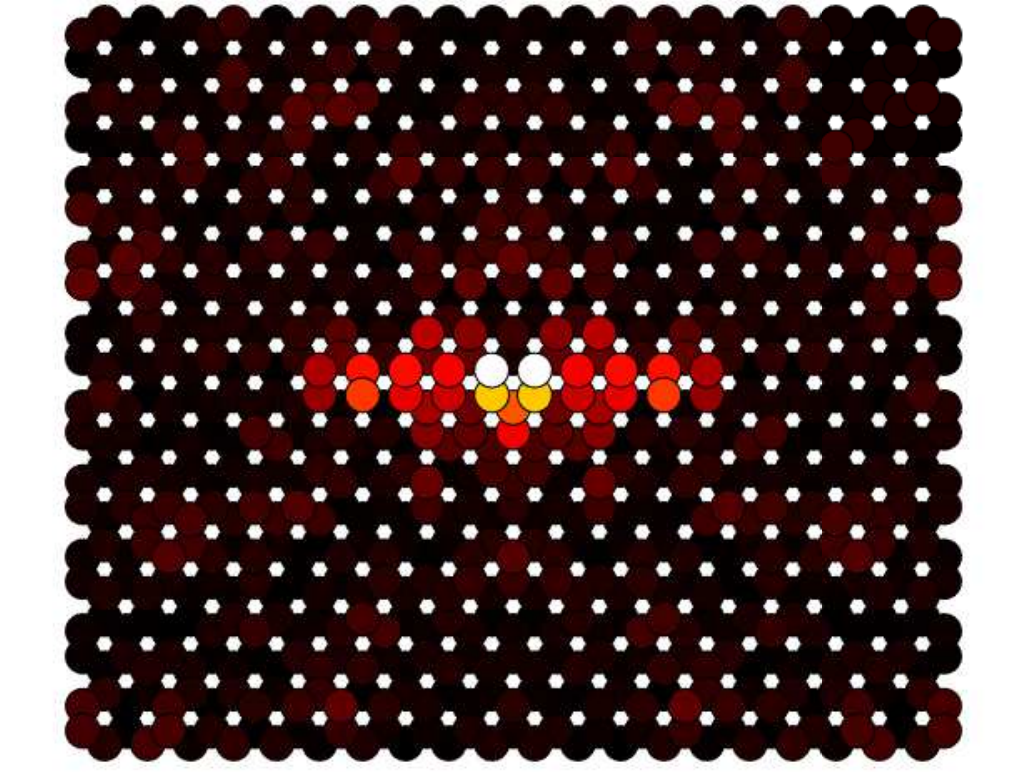} & \includegraphics[height=4.3cm, valign=c]{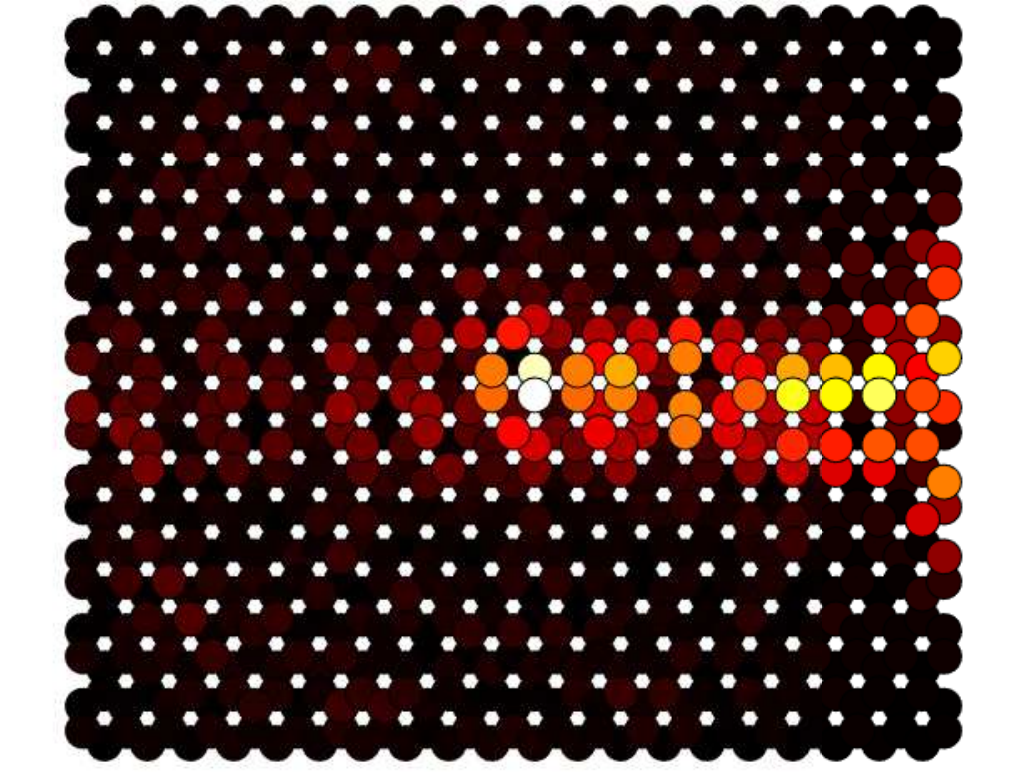} \\
\textbf{(c)} $f_2=1.38$ ${[kHz]}$ (blue arrow) & \textbf{(g)} $f_4=1.38$ ${[kHz]}$, $T_2$ & \textbf{(k)} $f_5=2.89$ ${[kHz]}$, $T_2$ \\
 \includegraphics[height=4.3cm, valign=c]{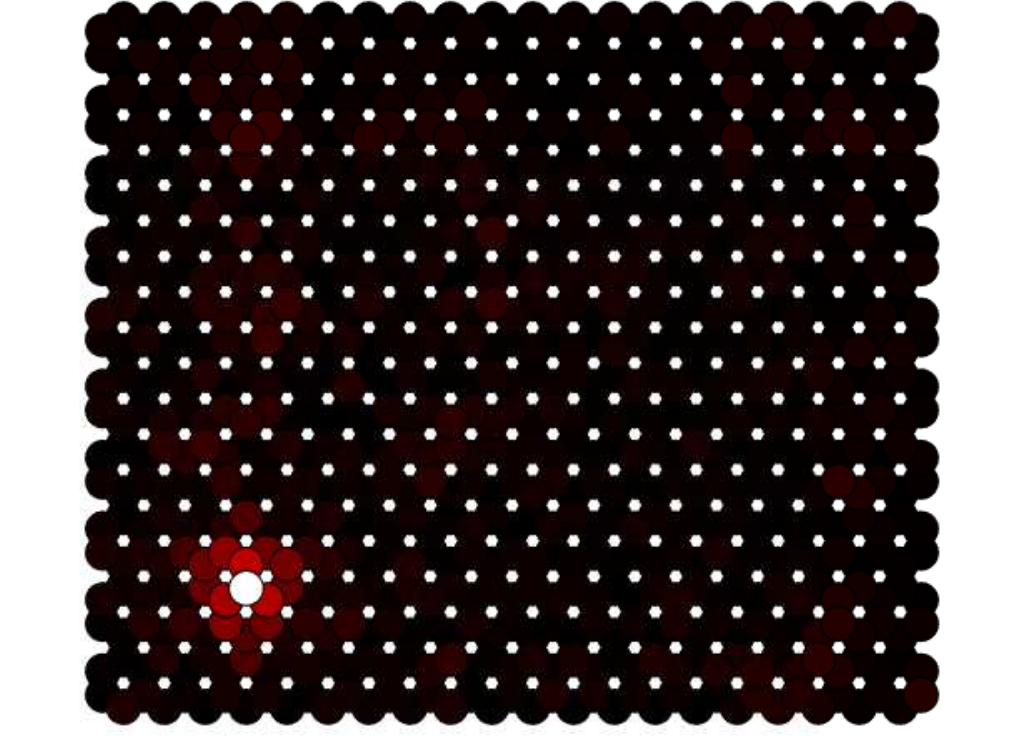} &  \includegraphics[height=4.3cm, valign=c]{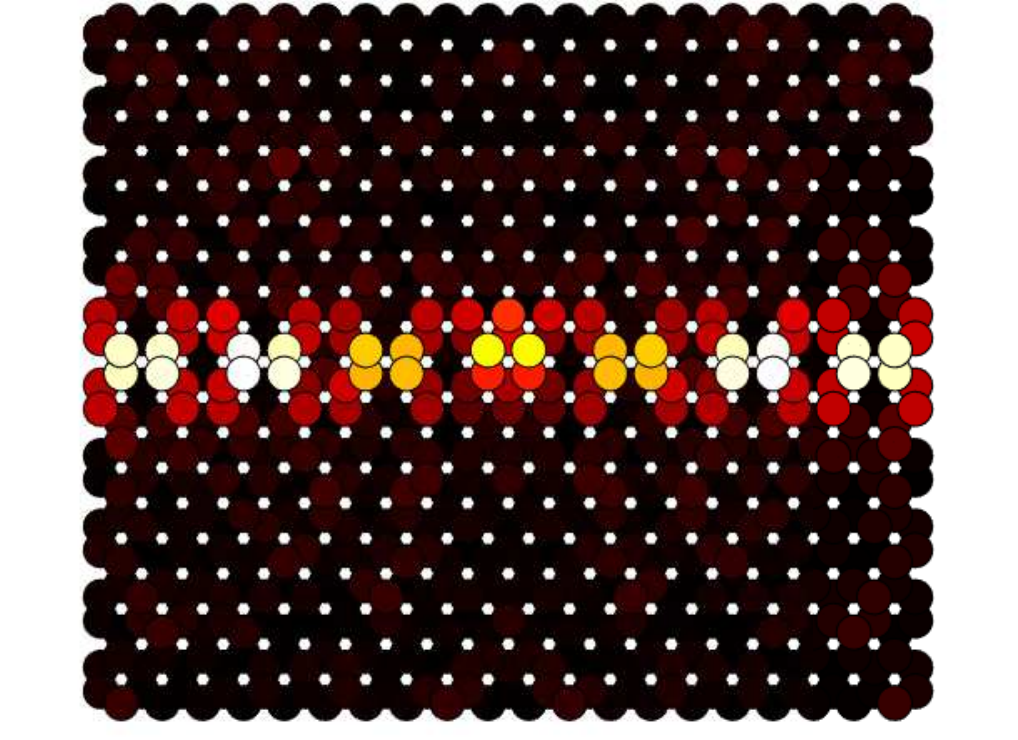} & \includegraphics[height=4.3cm, valign=c]{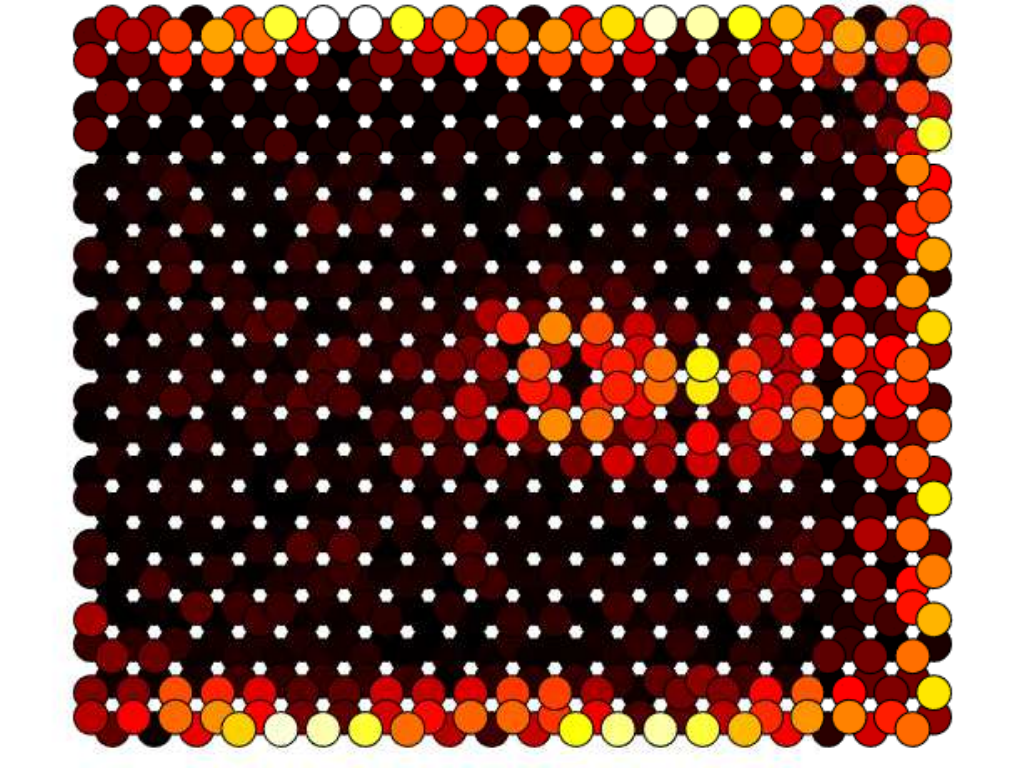} \\
\textbf{(d)}  $f_3=1$ ${[kHz]}$ (green arrow) & \textbf{(h)} Control signals & \textbf{(l)} Color scale \\
\includegraphics[height=4.3cm, valign=c]{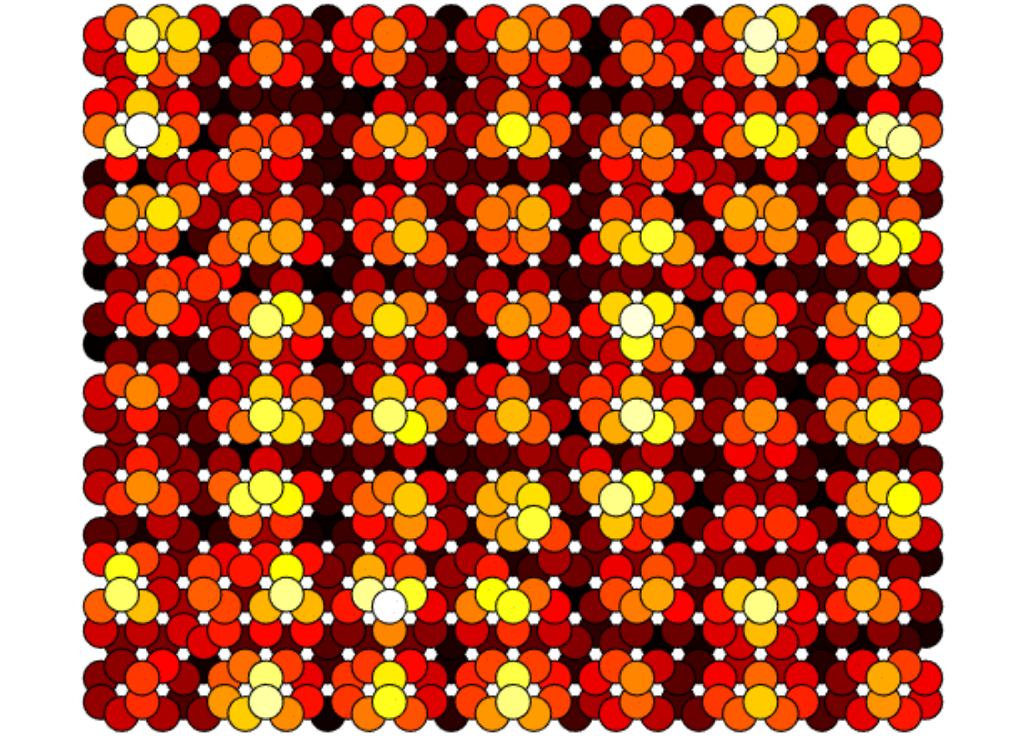} & \includegraphics[height=3.8cm, valign=c]{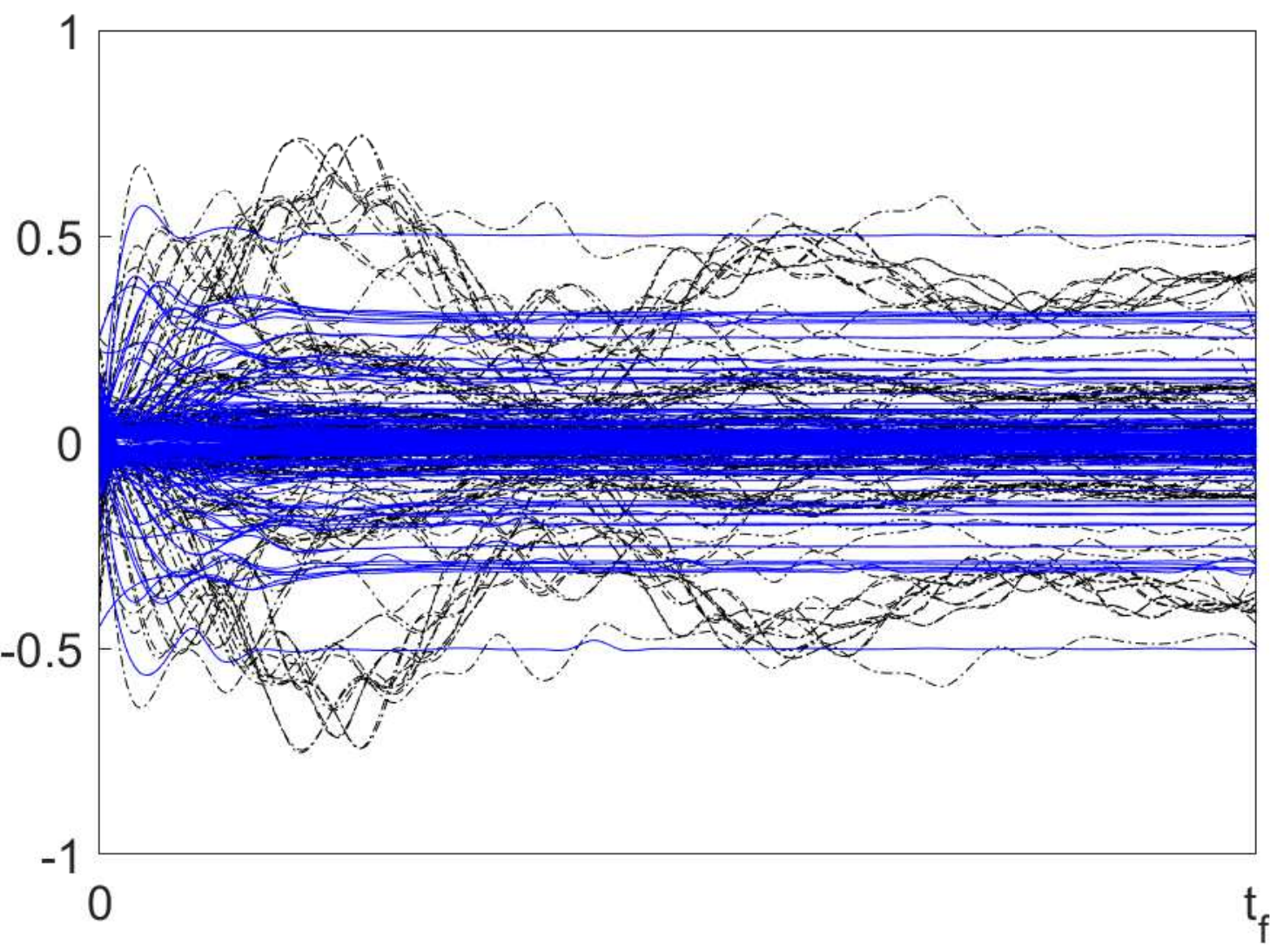} &  \includegraphics[height=3.0cm, valign=c]{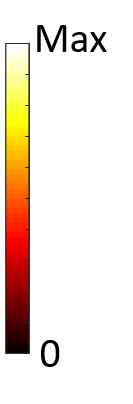}
\end{tabular}
\caption{Dynamical simulations of the autonomous acoustic metamaterial. (a) Control program 1: controllers $H_A(s)$ (black circles) and $H_B(s)$ (white circles) of \eqref{eq:Controller_H}, respectively create $Z_A(s)$ and $Z_B(s)$ impedances. (b) Time response to excitation on the interface (red arrow) at $f_1=1.38$ $[kHz]$. (c) Pressure field response to excitation in the bulk (blue arrow) at $f_2=1.38$ $[kHz]$. (d) Pressure field response to excitation in the bulk (green arrow) at $f_3=1$ $[kHz]$. (e) Control program 2: the controllers of \eqref{eq:Cont_law} here create a straight line artificial interface (red line). (f) Pressure field response to excitation on the interface (red arrow) at $f_4=1.38$ $[kHz]$, with simulation run time $T_1$. (g) The same simulation as in (f) with simulation run time $T_2>T_1$. {(h) Control signal envelopes for damping $D_h=0.01$ (black) and $D_h=0.1$ (blue). (i) Control program 3: the control law in \eqref{eq:Cont_law_QHE} creates directional couplings that break time reversal symmetry, and artificially divides the waveguide (red line) into a clockwise and counterclockwise polarization. (j) Pressure field response to excitation on the interface (red arrow) at $f_5=2.89$ $[kHz]$, with simulation run time $T_1$. (k) The same simulation as in (j) with simulation run time $T_2>T_1$.} (l) Color scale for (b-d),(f-g),(j-k), per figure.}
\label{Time_domain_sim}
\end{figure}

In this section we demonstrate that the feedback-based design of Sec. \ref{Setup} indeed converts a slab waveguide into an acoustic metamaterial, which supports steering of topologically protected, curved sound beams. First we consider the QVHE example of Sec. \ref{Target_gen}-\ref{Target}, and design control programs 1 and 2, respectively corresponding to the left and middle column in Fig.  \ref{Time_domain_sim}.
We perform dynamical simulations of the equivalent discrete system of Sec. \ref{Edge_disp}, whose dispersion model was fitted in Appendix \ref{Open} to the original hybrid continuous-discrete metamaterial.  
The simulated system corresponds to the slab waveguide in Fig. \ref{Scheme}(a)-(c). It is finite-sized in both $x$ and $y$ axes, and contains $10\times 20$ honeycomb cells.
To demonstrate the versatility of the underlying control mechanism, we program the metamaterial to generate topological interfaces of identical adjacent impedance, of two different shapes. The first interface is created by control program 1, and is $Z$-shaped, as illustrated in Fig. \ref{Time_domain_sim}(a). Each circle corresponds to an acoustic transducer that is driven in a real-time closed loop. The black and white fillings respectively indicate the controllers $H_A(s)$ and $H_B(s)$ in \eqref{eq:Controller_H}, which create the closed-loop impedances $Z_A(s)$ and $Z_B(s)$ in \eqref{eq:Z_12}. 
We stress that there is no physical interface between the waveguide plates, and all the actuators are identical. %
The values used for the simulation are those that were used in Sec. \ref{Inf_an} and \ref{Edge_disp} with the addition of small damping $D_h=0.01$ to the target impedance. \\
We perform three different simulations, each of them for a different combination of the source input location and frequency. The time duration of all the simulations is $t_f=0.5$ $[sec]$. 
The first simulation comprises a source at a location indicated by the red arrow, operated at the frequency $f_1=1.38$ $[kHz]$.
The resulting time response is plotted in Fig. \ref{Time_domain_sim}(b).
Since the interface state in Fig. \ref{Bandstructure_edge}(b) lies within the bulk band gap, waves propagate along the interface only.
Due to topological protection, implied by the topological invariant that is calculated in Appendix \ref{Top_inv}, the waves are completely immune to back-scattering from the sharp corners of the $Z$ shaped interface, and remain localized on the interface while smoothly traversing the corners.
Reflection from the metamaterial boundaries at the $Z$ shape ends does take place, though, as these are not accounted in the semi-infinite dispersion in Fig. \ref{Bandstructure_edge}(b). 
As for the control effort required for the closed-loop operation, the highest control inputs amplitude of the acoustic actuators was recorded on the interface and at its primary vicinity, reaching four times the source amplitude. The control effort can be reduced by reducing the band gap, which, in turn, will reduce the wave decay length outside the interface. \\
The second simulation comprises a source of the same frequency $f_2=1.38$ $[kHz]$, but is located away of the interface, as indicated by the blue arrow. The resulting time response is plotted in Fig. \ref{Time_domain_sim}(c).
Since the frequency falls within the bulk band gap, and the source is located at the bulk far from the interface, no wave propagation takes place.
In the third simulation the source loudspeaker is located at the same position as in the first case, as indicated by the green arrow, and operates at the frequency $f_3=1$ $[kHz]$.
The resulting time response is plotted in Fig. \ref{Time_domain_sim}(d).
Since $f_3$ falls within the bulk states of the dispersion relation, as shown in Fig. \ref{Bandstructure_edge}(b), waves are propagating everywhere along the metamaterial, and are reflecting from its boundaries back and forth (which is true regardless of the source location). \\
The interface shape and orientation are determined exclusively by the control program, and can be rearranged at will. 
{In Fig. \ref{Time_domain_sim}(e) we consider the same physical platform with control program 2 to create an artificial interface in a form of a straight line. 
We perform two simulations, in which the system is excited on this new interface (red arrow in (e)) at frequency $f_4=1.38$ $[kHz]$. The resulting time response is depicted in Fig. \ref{Time_domain_sim}(f) and \ref{Time_domain_sim}(g) at two time instances, $T_2>T_1$. We obtain a topologically protected sound beam propagation along the straight line interface. As expected for the QVHE, the propagation takes place in both directions from the excitation point. 
In Fig. \ref{Time_domain_sim}(h) we plot the envelopes (normalized by the source signal amplitude) of the corresponding control signals, i.e. the time responses of the acoustic actuators in \eqref{eq:Controller_H}, for two different values of target resonators damping, $D_h=0.01$ (black) and $D_h=0.1$ (blue). One observes that the increase in damping leads to a lower amplitude and a faster convergence of the control signals, traded-off with the pressure response amplitudes (not shown).} \\
{In the right column we demonstrate guiding of uni-directional sound beams (mimicking the quantum Hall effect) using the strategy of Sec. \ref{Target_QHE}, which we denote here by control program 3. We apply the controller in \eqref{eq:Cont_law_QHE} with the parameters $B_0$, $M_0$ and $\eta$ of Sec. \ref{Bandstructure_edge} (corresponding to the equivalent model of Sec. \ref{Open}), and with the gain parameter $\beta=0.2$. The parameter $\phi$ is set to $+\pi/3$ for actuators in one half of the waveguide, and to $-\pi/3$ for actuators in its second half, thus creating an artificial real-time interface, as illustrated by the red line if Fig. \ref{Time_domain_sim}(i). The $A-A$ and $B-B$ site feedback couplings through the interface are turned off. Since a counterclockwise propagation of beams is supported along the perimeter of the region with a positive phase $\phi$, and a clockwise propagation for a negative one, on the interface the beams can propagate only to the right. This is indeed obtained in closed-loop when we excite the metamaterial at the middle of the interface (red arrow), as demonstrated in Fig. \ref{Time_domain_sim}(j) at time instance $T_1$. 
When this uni-directional beam reaches the edge, it splits into two beams, which continue to propagate along the edges in the supported direction. This is completely different from the dynamical behavior of the QVHE case, for which the beams (e.g. in Fig. \ref{Time_domain_sim}(g)) that reach the edges, reflect from them back into the interface. 
Due to topological protection guaranteed by the nontrivial topological characteristic of the closed-loop dispersion relation (created by the control program), these uni-directional beams smoothly circumvent the sharp corners of the slab without any back-scattering.}

\section{Conclusion}   \label{Summary}

We presented a method to design acoustic metamaterials supporting propagation of sound beams of arbitrary reconfigurable shapes, in two-dimensional free space.
The underlying platform is a slab waveguide with an embedded feedback control mechanism, 
which enables shaping the sound pressure field between the plates in real-time, in a way that mimics quantum topological wave phenomena. 
Specifically, the model includes identical acoustic actuators mounted in a periodic pattern in one of the plates, and operated according to measurements from a mirror pattern of acoustic sensors, which are processed through autonomous electronic controllers. \\
As an example, we programmed the metamaterial to mimic the QVHE,
using a theoretical model that we developed for the closed-loop system. The required spatial symmetry breaking was created by a collocated pressure feedback at each lattice site,
augmenting the continuous pressure field by a discrete pattern of alternating acoustic impedances. 
We then demonstrated that the closed-loop metamaterial obtained a topological dispersion profile corresponding to the QVHE. 
We used numerical simulations to demonstrate the associated topological wave propagation. 
For a source of frequency in the bulk bandgap, located near the interface, robust, back-scattering-immune sound waves propagated between the plates, perfectly aligned with the interface. This is although no physical interface was present, and the space between the plates remained completely free. 
By reprogramming the embedded controllers, we realized trajectories of two different shapes. \\
Since the particular couplings and the consequent dynamical properties are exclusively defined by the algorithm
that is programmed into the controller, the feedback-based metamaterial is able to sustain any couplings and any
dynamical properties, within hardware limits and system stability.
{We demonstrated this versatility by deriving a different control algorithm, involving a non-collocated pressure feedback, which turned the topologically protected sound beams into uni-directional.}

\begin{ack}                             
We thank Moshe Goldstein for fruitful discussions. This research was supported in part by the Israel Science Foundation Grants No. 968/16 and 2096/18, by the Israeli Ministry of Science and Technology Grant No. 3-15671, by the US-Israel Binational Science Foundation Grant No. 2018226, and by the National Science Foundation Grant No. NSF PHY-1748958. YS thanks the Center for Nonlinear Studies at Los Alamos National Laboratory for its hospitality.
\end{ack}

\appendix

\section{Infinite system dispersion calculation using the Plane Wave Expansion method}  \label{PWE}   

Here we present a detailed frequency dispersion calculation of the system in \eqref{eq:res_2D_2}, which models the acoustic metamaterial in Fig. \ref{Scheme}(b),(c), when operated in closed-loop.
For the sake of the calculation according to Bloch theory of periodic systems \cite{wilcox1977theory}, in this section the metamaterial is assumed of infinite extension. 
The calculation can be therefore folded into a single unit cell.
Following the unit cell geometry in Fig. \ref{Bandstructure_infinite}(b)-left, one obtains $b=a/\sqrt{3}$ (where $a$ is the lattice constant and $b$ is the distance between the $A$ and $B$ sites), as well as $R_{A1}=R_{A2}=b\sin{30^o}/\sin{120^o}/2$, leading to $R_{A1}=R_{A2}=a/6$.
Since this system is a hybridization of the continuous pressure field $p(\textbf{r};s)$ and the discretely located transducers, we can calculate its dispersion relation using, for example, the Plane Wave Expansion method \cite{chaunsali2018subwavelength,xiao2012flexural,wang2015tuning}. 
This method assumes a series solution of traveling harmonic waves, $p(\textbf{r},t)=e^{i\omega t}P(\textbf{r})$, where
\begin{equation}  \label{eq:PWE_sol}
    P(\textbf{r})=\sum_{m,n=-M}^Mp_\textbf{G}e^{-i(\textbf{k}+\textbf{G})\cdot \textbf{r}}.
\end{equation}
Here $\textbf{r}=r_1\textbf{d}_1+r_2\textbf{d}_2$ is the position vector in the real lattice space, and $\textbf{k}=k_1\textbf{b}_1+k_2\textbf{b}_2$ is the base wavevector in the reciprocal lattice space, defined by
\begin{equation} \label{eq:d1d2b1b2}
    \textbf{d}_1=(a,0), \quad \textbf{d}_2=\left(a/2,\sqrt{3}a/2\right), \quad \textbf{b}_1=2\pi/a\left(1,-1/\sqrt{3}\right), \quad \textbf{b}_2=2\pi/a\left(0,2/\sqrt{3}\right).
\end{equation}
$\textbf{G}=m\textbf{b}_1+n\textbf{b}_2$ is its expansion, where $m$ and $n$ are integer indices, which span across $-M:M$, and $M$ is the series truncation order. 
Since $\textbf{G}$ covers a two-dimensional grid, the total number of its entries is $N^2$, where $N=2M+1$.
Since $p(\textbf{r};s)$ transforms to frequency domain when $s=i\omega$, we substitute \eqref{eq:PWE_sol} in \eqref{eq:res_2D_2}, which reads
\begin{equation}  \label{eq:PWE_sol_subs}
    \frac{c^2}{\omega^2}\sum_{n,m=-M}^M|\textbf{k}+\textbf{G}|^2e^{-i(\textbf{k}+\textbf{G})\cdot \textbf{r}}p_\textbf{G}=-\sum_{n,m=-M}^M\left\{e^{-i(\textbf{k}+\textbf{G})\cdot \textbf{r}}+\rho c^2\eta \sum_{j=A,B} \frac{e^{-i(\textbf{k}+\textbf{G})\cdot \textbf{R}_j}}{\widetilde{Z}_j(\omega)} \delta(\textbf{r}-\textbf{R}_j)\right\}p_\textbf{G},
\end{equation}
where $\widetilde{Z}_{A,B}(\omega)= i\omega Z_{A,B}(i\omega)=-M_{A,B}\omega^2+K_{A,B}$ (for the sake of the calculation we set $D_h\rightarrow 0$). Multiplying \eqref{eq:PWE_sol_subs} by $e^{i(\textbf{k}+\widehat{\textbf{G}})\cdot \textbf{r}}$, where $\widehat{\textbf{G}}=\widehat{m}\textbf{b}_1+\widehat{n}\textbf{b}_2$ for some specific values $\widehat{m}$ and $\widehat{n}$, we obtain
\begin{equation}  \label{eq:PWE_sol_orth}
    \frac{c^2}{\omega^2}\sum_{n,m=-M}^M|\textbf{k}+\textbf{G}|^2e^{-i(\textbf{G}-\widehat{\textbf{G}})\cdot \textbf{r}}p_\textbf{G}=-\sum_{n,m=-M}^M\left\{e^{-i(\textbf{G}-\widehat{\textbf{G}})\cdot \textbf{r}}+\rho c^2\eta e^{i(\textbf{k}+\widehat{\textbf{G}})\cdot \textbf{r}}\sum_{j=A,B} \frac{e^{-i(\textbf{k}+\textbf{G})\cdot \textbf{R}_j}}{\widetilde{Z}_j(\omega)} \delta(\textbf{r}-\textbf{R}_j)\right\}p_\textbf{G}.
\end{equation}
Due to the orthogonality property of the Fourier series, we have
\begin{equation}  \label{eq:PWE_int_terms}
    \iint_{A_c}e^{-i(\textbf{G}-\widehat{\textbf{G}})\cdot \textbf{r}}\mathrm{d}A_c=\begin{cases}A_c, \quad  &\textbf{G}=\widehat{\textbf{G}} \\ 0, \quad &\textbf{G}\neq\widehat{\textbf{G}}\end{cases} \qquad , \qquad \iint_{A_c}f(\textbf{r})\delta(\textbf{r}-\textbf{R}_\alpha)\mathrm{d}A_c=f(\textbf{R}_\alpha),
\end{equation}
where $A_c=a^2/\sqrt{3}$ is the unit cell area (Fig. \ref{Scheme}(b)), and $a$ is the lattice constant. Integrating \eqref{eq:PWE_sol_orth} over a unit cell then gives
\begin{equation}  \label{eq:PWE_sol_int}
    \frac{c^2}{\omega^2}|\textbf{k}+\widehat{\textbf{G}}|^2A_c p_{\widehat{\textbf{G}}}=-A_c p_{\widehat{\textbf{G}}} -\rho c^2\eta \sum_{m,n=-M}^M \sum_{j=A,B}\frac{e^{-i(\textbf{G}+\widehat{\textbf{G}})\cdot \textbf{R}_j}}{\widetilde{Z}_j(\omega)}p_\textbf{G}.
\end{equation}
Using matrix formulation, we define $\sum_{m,n=-M}^M e^{-i(\textbf{G}+\widehat{\textbf{G}})\cdot \textbf{R}_j}p_\textbf{G}=E_j p_{\widehat{\textbf{G}}}$, where
\begin{equation}  \label{eq:E_j}
    E_j=e^{i\left[\begin{array}{cccc}{\textbf{G}_1}
         & \textbf{G}_2 & \cdots & \textbf{G}_{N^2}\end{array}\right]^T\cdot \textbf{R}_j}\cdot e^{i\textbf{R}_j\cdot\left[\begin{array}{cccc}{\textbf{G}_1}
         & \textbf{G}_2 & \cdots & \textbf{G}_{N^2}\end{array}\right]}.
\end{equation}
Following the target resonators coordinates that are given in \eqref{eq:d1d2b1b2}, we obtain $\textbf{G}\cdot\textbf{R}_A=2\pi a(m+n)/6$, and $\textbf{G}\cdot\textbf{R}_B=-2\pi a(m+n)/6$.
Substituting \eqref{eq:E_j} into \eqref{eq:PWE_sol_int} and using the explicit form of $\widetilde{Z}_j(\omega)$, \eqref{eq:PWE_sol_int} takes the form
\begin{equation}  \label{eq:PWE_sol_int_exp}
    [M_AM_B\omega^4-(M_AK_B+M_BK_A)\omega^2+K_AK_B]\left(c^2|\textbf{k}+\widehat{\textbf{G}}|^2A_c-\omega^2\right)p_{\widehat{\textbf{G}}}=-\rho c^2\eta\omega^2 \left[(M_B\omega^2+K_B)E_A+(M_A\omega^2+K_A)E_B\right]p_{\widehat{\textbf{G}}},
\end{equation}
where $E_j$ is defined in \eqref{eq:E_j}. Rearranging \eqref{eq:PWE_sol_int_exp} results in the following polynomial eigenvalue problem,
\begin{equation}  \label{eq:poly_eig}
    \left(\textbf{A}_3\lambda^3+\textbf{A}_2\lambda^2+\textbf{A}_1\lambda+\textbf{A}_0\right)p_{\widehat{\textbf{G}}}=0 \qquad , \qquad \lambda=\omega^2,
\end{equation}
where
\begin{equation}  \label{eq:poly_eig_A_3210}
\begin{alignedat}{2}
 \textbf{A}_3&=M_AM_B\textbf{A}, \quad &&\textbf{A}_2=M_AM_Bc^2\textbf{B}-(M_AK_B+M_BK_A)\textbf{A}-\rho c^2\eta\textbf{C}_M, \\ \textbf{A}_1&=-(M_AK_B+M_BK_A)c^2\textbf{B}+K_AK_B\textbf{A}+\rho c^2\eta\textbf{C}_K, \quad &&\textbf{A}_0=K_AK_Bc^2\textbf{B},
\end{alignedat}
\end{equation}
and
\begin{equation}  \label{eq:poly_eig_ABC_MC_K}
    \textbf{A}=A_c\textbf{I}_{N^2}, \quad \textbf{B}=-A_c\left(\begin{array}{cc cc} |\textbf{k}+\widehat{\textbf{G}}_1|^2 &  & & \\
         & |\textbf{k}+\widehat{\textbf{G}}_2|^2 & & \\
         & & \cdots & \\
         & & & |\textbf{k}+\widehat{\textbf{G}}_{N^2}|^2
    \end{array}\right), \quad \textbf{C}_M=M_AE_A+M_BE_B, \quad \textbf{C}_K=K_AE_A+K_BE_B.
\end{equation}
We then rewrite \eqref{eq:poly_eig}-\eqref{eq:poly_eig_ABC_MC_K} in a companion form to obtain an augmented linear eigenvalue problem
\begin{equation}  \label{eq:lin_eig}
    \lambda\textbf{P}\textbf{v}=\textbf{Q}\textbf{v},
\end{equation}
where
\begin{equation}  \label{eq:lin_eig_mat}
    \textbf{P}=\left(\begin{array}{ccc} \textbf{I} & 0 & 0\\
       0  & \textbf{I} & 0\\
        0 & 0 & \textbf{A}_3
    \end{array}\right) \qquad , \qquad \textbf{Q}=\left(\begin{array}{ccc} 0 & \textbf{I} & 0 \\
        0 & 0 & \textbf{I} \\
        -\textbf{A}_0 & -\textbf{A}_1 & \textbf{A}_2
    \end{array}\right),
\end{equation}
and $\textbf{v}$ is the augmented eigenvector of length $3N^2$.
The dispersion relations of the infinite closed-loop system that are shown in Fig. \ref{Bandstructure_infinite} are the first two solutions of \eqref{eq:lin_eig}-\eqref{eq:lin_eig_mat} for $N=4$.

\section{The topological invariant calculation}   \label{Top_inv}

\begin{figure}[tpb]
\centering
\begin{tabular}{l l l l}
\textbf{(a)}  & \includegraphics[height=4.8cm, valign=c]{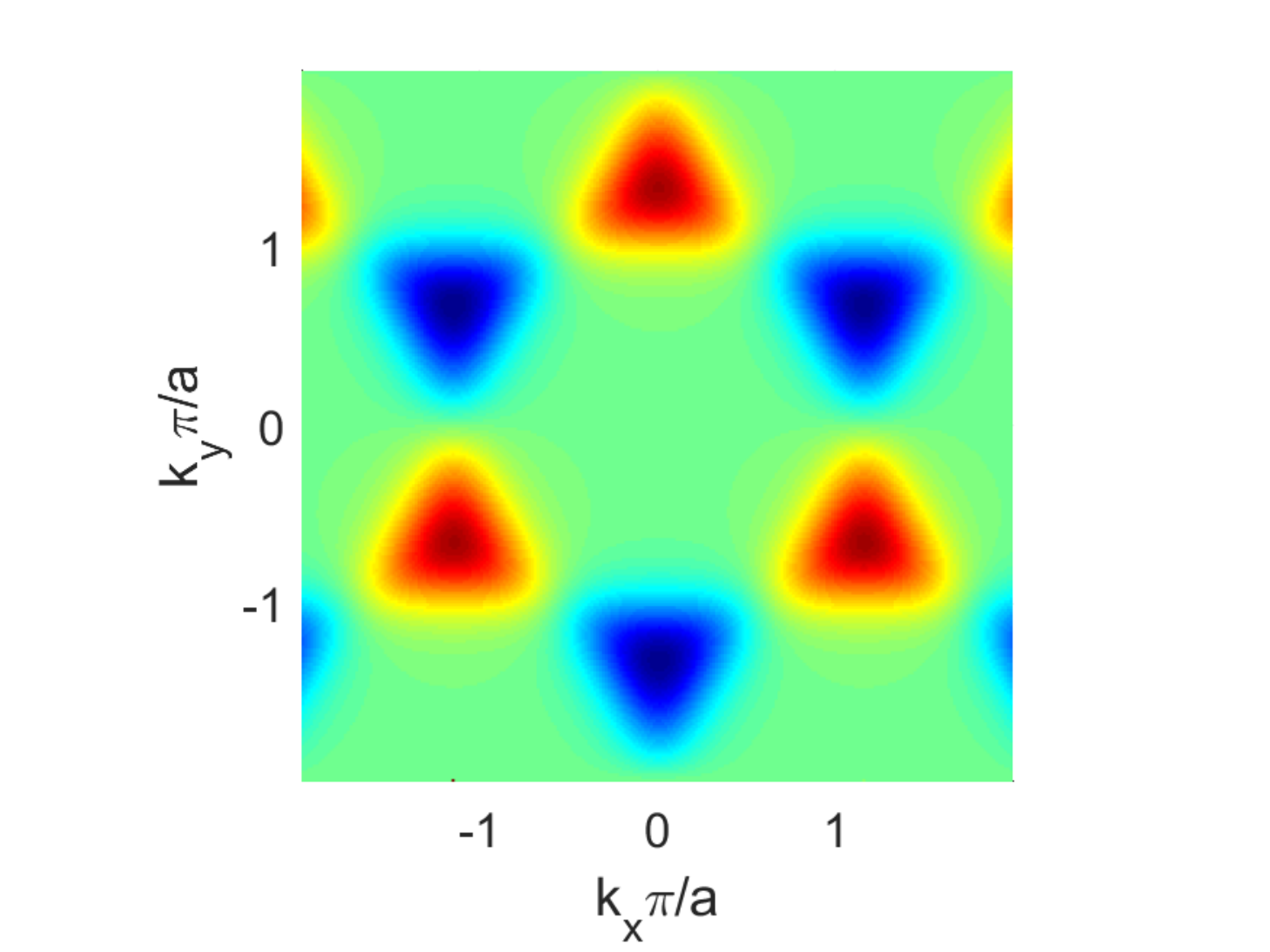}     &  \textbf{(b)} & \includegraphics[height=4.0cm, valign= c]{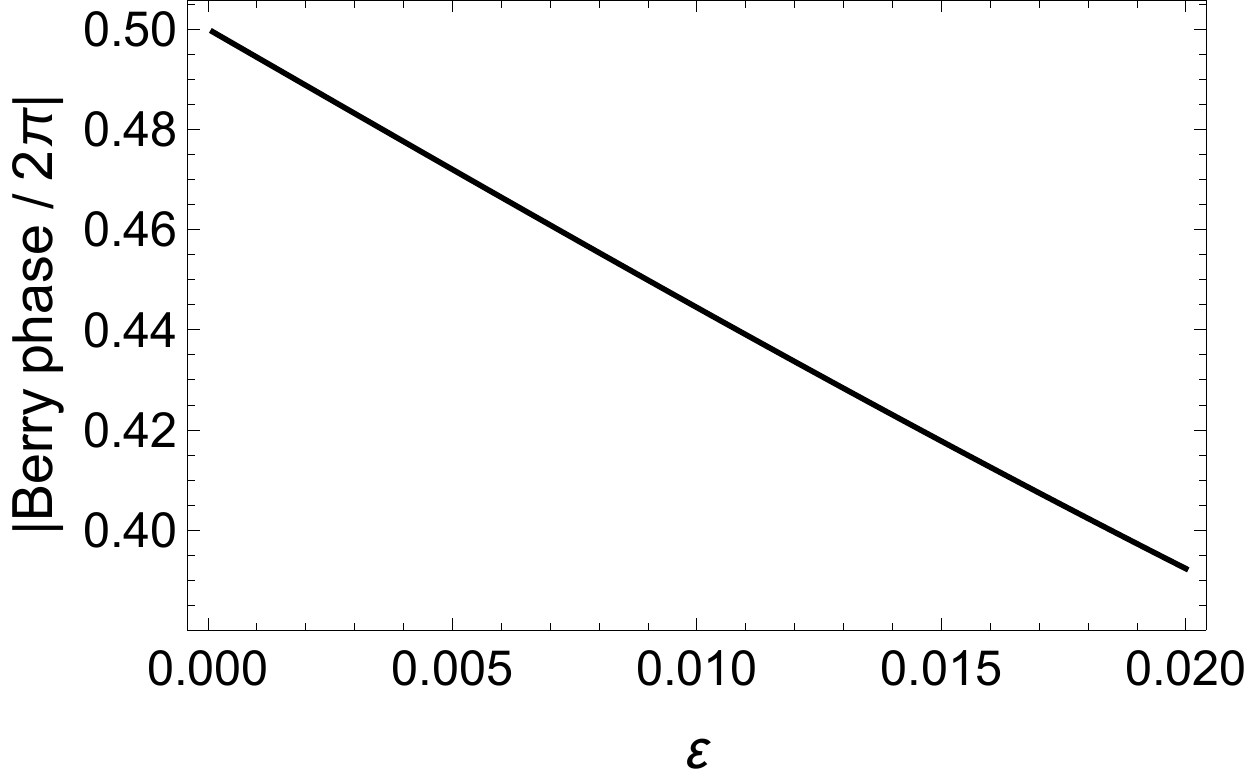}
\end{tabular}
\caption{(a) Contour plot of the Berry curvature in the first Brillouin zone, for $\epsilon=0.1$. The non-zero curvature values are centered around the $K$ and $K'$ high symmetry points of Fig. \ref{Bandstructure_infinite}(a)-right. (b) The Chern number $|C_K|$ is linearly approaching $1/2$ with $\epsilon\rightarrow 0$, as expected from equation \eqref{eq:ChernvsGap}.}
\label{fig:ChernComputation}
\end{figure}

The topological character of systems supporting the QVHE (and its classical analogies) manifests itself through the valley Chern number \cite{franz2013topological}, evaluated for the gapped bands of the infinite system dispersion profile.
In our system the relevant bands are the two low frequency range solutions of the augmented eigenvalue problem in \eqref{eq:lin_eig}-\eqref{eq:lin_eig_mat} for $\epsilon \neq 0$, depicted in Fig. \ref{Bandstructure_infinite}(e),(f). 
For each band, a Chern number is given by the formula \cite{franz2013topological}
\begin{equation}   \label{eq:Berry}
    C=\frac{1}{2\pi}\int_{BZ}\Omega_{\textbf{v}}\left(\textbf{k}\right)\textrm{d}\textbf{k}^2, \qquad \Omega_{\textbf{v}}\left(\textbf{k}\right)=\nabla_{\textbf{k}}\times(-i\textbf{v}(\textbf{k})^{\dagger}\nabla_{\textbf{k}}\textbf{v}(\textbf{k})),
\end{equation}
where $\Omega_{\textbf{v}}\left(\textbf{k}\right)$ is the Berry curvature \cite{franz2013topological}, $\textbf{v}\left(\textbf{k}\right)$ is the corresponding eigenstate and integration is performed over the entire Brillouin zone. As a consequence of time reversal symmetry, the Chern number in our system is zero, but a different topological index, the valley Chern number, results in a finite quantized value. 
The valley Chern number is defined as $C_V=C_K-C_{K'}$, where $C_K$ and $C_{K'}$ are computed by integrating the Berry curvature over the infinite wavevector space of the linearized low frequency model around the high symmetry points $K$ and $K'$, as illustrated in Fig.~\ref{fig:ChernComputation}(a).\\
Employing the numerically optimized algorithm \cite{Fukui2005} in the $\epsilon>0$ case, we obtain a Berry curvature comprised of alternating vortexes at the $K$ and $K'$ points, as illustrated in Fig.~\ref{fig:ChernComputation}(a).
We integrate the Berry curvature over a small region of radius $R_{\textbf{k}}$ around $K$ or $K'$, and find that the result (divided by $2\pi$) converges to $\pm1/2$ as the band-gap between the two lowest bands, $\Delta(\epsilon)$, gets smaller, and as $R_{\textbf{k}}$ gets larger. $\Delta(\epsilon)$ increases with $|\epsilon|$, and their exact functional dependence can be found by linearizing the eigenvalue problem in \eqref{eq:lin_eig}-\eqref{eq:lin_eig_mat} around $K$ or $K'$, where the gap size is controlled by $\epsilon$, and deriving an effective low frequency model, where $\Delta$ directly describes the gap. Here we look at a limit of a small $\epsilon$, where we can assume $\Delta(\epsilon)\propto\epsilon$, and in Fig.~\ref{fig:ChernComputation}(b) we plot the Berry phase around $K$ as a function of $\epsilon$, finding a linear dependence consistent with
\begin{equation}
    \frac{1}{2\pi}\int_{R_{K}}\Omega_{\textbf{v}}\left(\textbf{k}\right)\textrm{d}\textbf{k}^2=\pm\frac{1}{2}\left(1-\alpha\frac{\Delta(\epsilon)}{R_{\textbf{k}}}\right).
    \label{eq:ChernvsGap}
\end{equation}
The pre-factor $\alpha$ results from the linearization and accounts for units, as $\Delta(\epsilon)$ is an effective gap term in units of frequency, $R_k$ is in units of $1/a$, and the Berry phase is unit-less. While equation \eqref{eq:ChernvsGap} was numerically deduced from the model, where $R_{\textbf{k}}$ can be increased up to a finite limit for the integration to encompass solely a single Dirac point, it also describes the linearized low energy model around the $K$ and $K'$ points. For the linearized models, the limit $R_K\rightarrow\infty$ leads to, as prescribed by equation \eqref{eq:Berry}, $C_K,C_{K'}\rightarrow\pm 1/2$, resulting in a non-vanishing valley Chern number and an associated topologically non-trivial phase, analogous to the QVHE. {We note that when $\epsilon$ flips sign, the Berry curvature in Fig.~\ref{fig:ChernComputation}(a) obtains the mirror image through the $k_x-k_y$ plane, and the values of $C_K$ and $C_{K'}$ flip sign as well, corresponding to the topological phase transition discussed in Sec. \ref{Inf_an}.}

\section{Equivalent discrete model derivation} \label{Open}

\begin{figure}[tbp]
\centering
\begin{tabular}{l l l l}
\textbf{(a)} & 
\includegraphics[width=4.2cm, valign=t]{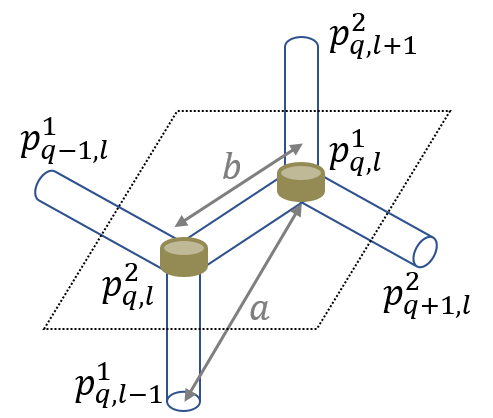} & \textbf{(b)}  &
\includegraphics[width=6.0cm, valign=t]{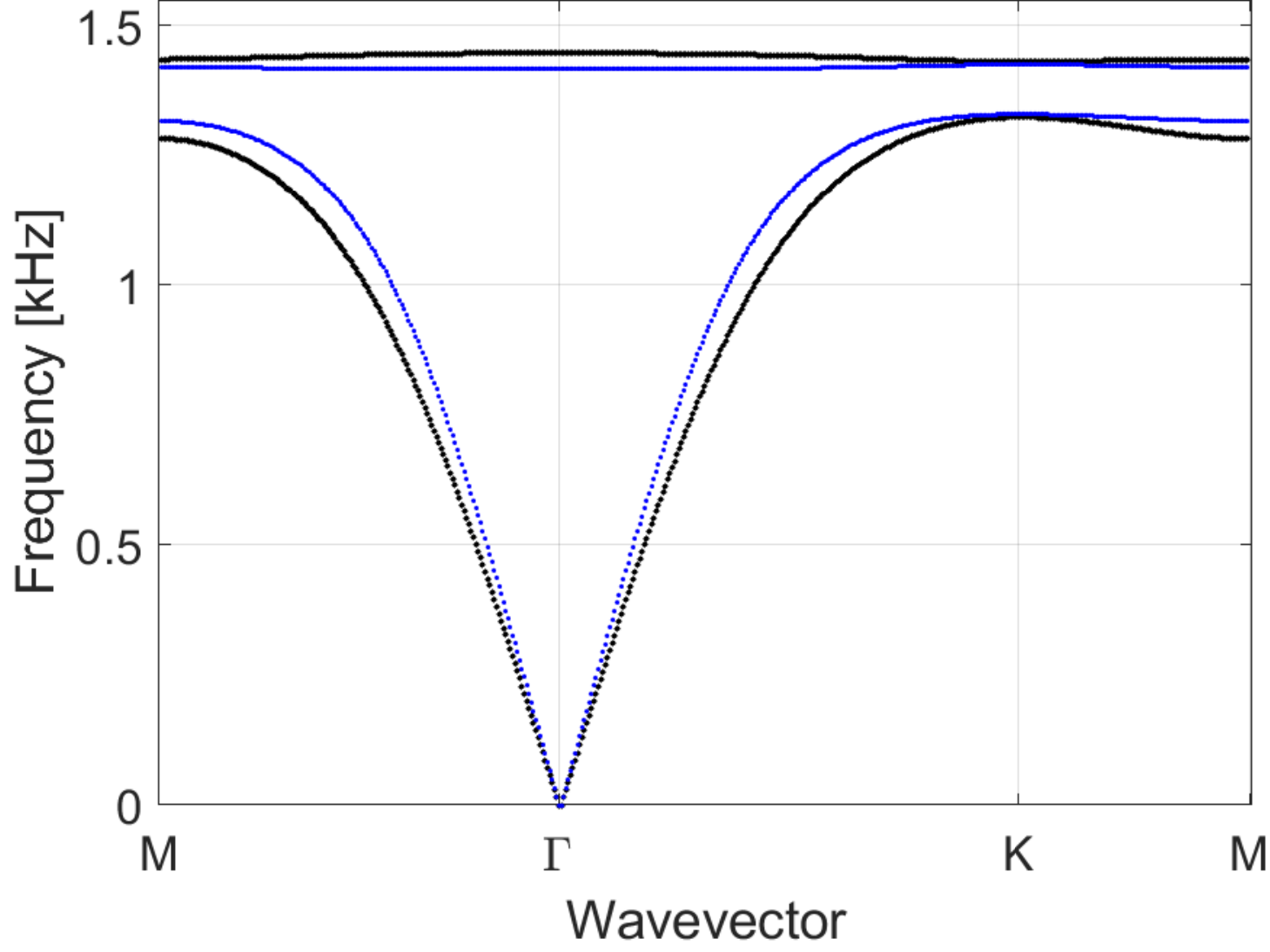}
\end{tabular}
\caption{\small Equivalent discrete model of the hybrid continuous-discrete system in \eqref{eq:res_2D_2}. (a) Schematic of a waveguide unit cell with air flow reduced to direct paths between transducers. (b) Fitting of the infinite-periodic systems dispersion relations (black - original, blue - equivalent).}
\label{Eq_discrete}
\end{figure}

In this section we derive the equivalent discrete model of the hybrid continuous-discrete target system in \eqref{eq:res_2D_2}, in order to calculate the dispersion relation of the open system in Sec. \ref{Edge_disp}. 
The only part of \eqref{eq:res_2D_2} to be discretized is the continuous two-dimensional wave equation on the left hand side. We thus completely preserve the coupling of the target resonators on the right hand side. 
We perform the discretization in two stages. First, we limit the sound pressure wave propagation to direct routes between the target resonators locations, through, for example, artificial one-dimensional tubes, as illustrated in Fig. \ref{Eq_discrete}(a).
Here $a$ is the same lattice constant as in the continuous case, and $b$ is the artificial tube length.
As a result, we obtain a periodic hexagonal net with two sites per unit cell, labeled by $A$ and $B$, as captured within the black parallelogram. 
The second stage is approximating the continuous pressure field in those artificial tubes by equivalent lumped acoustic springs. 
The governing target closed-loop equations of the $\{m,n\}$ unit cell for each of these sites, then respectively take the form
\begin{subequations}   \label{eq:Eq_model}
    \begin{align}
        B_0\left(p^B_{q,l+1}+p^B_{q,l}+p^B_{q+1,l}-3p^A_{q,l}\right)&=M_0s^2p^A_{q,l}+M_0B_0\widehat{\eta}\frac{s}{Z_A(s)}p^A_{q,l}, \\
        B_0\left(p^A_{q,l-1}+p^A_{q,l}+p^A_{q-1,l}-3p^B_{q,l}\right)&=M_0s^2p^B_{q,l}+M_0B_0\widehat{\eta}\frac{s}{Z_B(s)}p^B_{q,l},
    \end{align}
\end{subequations}
where $B_0=\rho c^2/b$ $[kg/s^2/m^2]$ and $M_0=\rho b$ $[kg/m^2]$ are the equivalent acoustic spring and mass of each artificial tube, respectively. The design parameter is given by $\widehat{\eta}=A_n/A_{\widehat{t}}$, and is dimensionless, as expected for a one-dimensional waveguide, where $A_n$ is the target resonator opening area (as before) and $A_{\widehat{t}}$ is the cross-section area of the artificial tube.\\
In order to fit the approximated model in \eqref{eq:Eq_model} to the exact one in \eqref{eq:res_2D_2}, we first calculate the dispersion relation of an infinite-sized approximated system. Transforming \eqref{eq:Eq_model} back into time domain, we obtain two fourth-order ordinary differential equations in time. Substituting then the traveling harmonic wave solution $\textbf{p}=\textbf{p}_0e^{i(k_1ma+k2_na-\omega t)}$, we obtain, similarly to Appendix \ref{PWE}, a polynomial eigenvalue problem, but this time it is quadratic. Finally, we arrive with the equivalent values for the target resonators parameter $Vol_{eq}=1.15Vol$, and the artificial tubes cross-section area $A_{\widehat{t}}=7^2A_n$. Keeping $L_{n\_eq}=L_n$, $A_{n\_eq}=A_n$ and $a_{eq}=a$, we obtain an exceptional fitting to the exact system dispersion relation, as illustrated in Fig. \ref{Eq_discrete}(b). Regarding then the system in \eqref{eq:Eq_model} as completely equivalent, we employ it to calculate the dispersion (band-structure) of a semi-infinite lattice in Sec. \ref{Edge_disp}, and for time domain simulations in Sec. \ref{Time_sim}.

\bibliographystyle{IEEEtran} 
\bibliography{AcousticQVHE}                      

\begin{thebibliography}{10}
\providecommand{\url}[1]{#1}
\csname url@samestyle\endcsname
\providecommand{\newblock}{\relax}
\providecommand{\bibinfo}[2]{#2}
\providecommand{\BIBentrySTDinterwordspacing}{\spaceskip=0pt\relax}
\providecommand{\BIBentryALTinterwordstretchfactor}{4}
\providecommand{\BIBentryALTinterwordspacing}{\spaceskip=\fontdimen2\font plus
\BIBentryALTinterwordstretchfactor\fontdimen3\font minus
  \fontdimen4\font\relax}
\providecommand{\BIBforeignlanguage}[2]{{%
\expandafter\ifx\csname l@#1\endcsname\relax
\typeout{** WARNING: IEEEtran.bst: No hyphenation pattern has been}%
\typeout{** loaded for the language `#1'. Using the pattern for}%
\typeout{** the default language instead.}%
\else
\language=\csname l@#1\endcsname
\fi
#2}}
\providecommand{\BIBdecl}{\relax}
\BIBdecl

\bibitem{shelby2001experimental}
R.~A. Shelby, D.~R. Smith, and S.~Schultz, ``Experimental verification of a
  negative index of refraction,'' \emph{Science}, vol. 292, no. 5514, pp.
  77--79, 2001.

\bibitem{cubukcu2003negative}
E.~Cubukcu, K.~Aydin, E.~Ozbay, S.~Foteinopoulou, and C.~M. Soukoulis,
  ``Negative refraction by photonic crystals,'' \emph{Nature}, vol. 423, no.
  6940, pp. 604--605, 2003.

\bibitem{pendry2000negative}
J.~B. Pendry, ``Negative refraction makes a perfect lens,'' \emph{Physical
  Review Letters}, vol.~85, no.~18, p. 3966, 2000.

\bibitem{schurig2006metamaterial}
D.~Schurig, J.~J. Mock, B.~Justice, S.~A. Cummer, J.~B. Pendry, A.~F. Starr,
  and D.~R. Smith, ``Metamaterial electromagnetic cloak at microwave
  frequencies,'' \emph{Science}, vol. 314, no. 5801, pp. 977--980, 2006.

\bibitem{ergin2010three}
T.~Ergin, N.~Stenger, P.~Brenner, J.~B. Pendry, and M.~Wegener,
  ``Three-dimensional invisibility cloak at optical wavelengths,''
  \emph{Science}, vol. 328, no. 5976, pp. 337--339, 2010.

\bibitem{soukoulis2011past}
C.~M. Soukoulis and M.~Wegener, ``Past achievements and future challenges in
  the development of three-dimensional photonic metamaterials,'' \emph{Nature
  Photonics}, vol.~5, no.~9, p. 523, 2011.

\bibitem{khelif2016phononic}
A.~Khelif and A.~Adibi, \emph{Phononic Crystals}.\hskip 1em plus 0.5em minus
  0.4em\relax Springer, 2016.

\bibitem{craster2012acoustic}
R.~V. Craster and S.~Guenneau, \emph{Acoustic metamaterials: {N}egative
  refraction, imaging, lensing and cloaking}.\hskip 1em plus 0.5em minus
  0.4em\relax Springer Science \& Business Media, 2012, vol. 166.

\bibitem{cummer2007one}
S.~A. Cummer and D.~Schurig, ``One path to acoustic cloaking,'' \emph{New
  Journal of Physics}, vol.~9, no.~3, p.~45, 2007.

\bibitem{seo2012acoustic}
Y.~M. Seo, J.~J. Park, S.~H. Lee, C.~M. Park, C.~K. Kim, and S.~H. Lee,
  ``Acoustic metamaterial exhibiting four different sign combinations of
  density and modulus,'' \emph{Journal of Applied Physics}, vol. 111, no.~2, p.
  023504, 2012.

\bibitem{dubois2017observation}
M.~Dubois, C.~Shi, X.~Zhu, Y.~Wang, and X.~Zhang, ``Observation of acoustic
  dirac-like cone and double zero refractive index,'' \emph{Nature
  Communications}, vol.~8, p. 14871, 2017.

\bibitem{rohde2015experimental}
C.~A. Rohde, T.~P. Martin, M.~D. Guild, C.~N. Layman, C.~J. Naify, M.~Nicholas,
  A.~L. Thangawng, D.~C. Calvo, and G.~J. Orris, ``Experimental demonstration
  of underwater acoustic scattering cancellation,'' \emph{Scientific Reports},
  vol.~5, p. srep13175, 2015.

\bibitem{zhu2011holey}
J.~Zhu, J.~Christensen, J.~Jung, L.~Martin-Moreno, X.~Yin, L.~Fok, X.~Zhang,
  and F.~Garcia-Vidal, ``A holey-structured metamaterial for acoustic
  deep-subwavelength imaging,'' \emph{Nature Physics}, vol.~7, no.~1, pp.
  52--55, 2011.

\bibitem{cummer2016controlling}
S.~A. Cummer, J.~Christensen, and A.~Al{\`u}, ``Controlling sound with acoustic
  metamaterials,'' \emph{Nature Reviews Materials}, vol.~1, no.~3, p. 16001,
  2016.

\bibitem{liu2011elastic}
X.-N. Liu, G.-K. Hu, G.-L. Huang, and C.-T. Sun, ``An elastic metamaterial with
  simultaneously negative mass density and bulk modulus,'' \emph{Applied
  Physics Letters}, vol.~98, no.~25, p. 251907, 2011.

\bibitem{hu146metamaterial}
G.~Hu, A.~C. Austin, V.~Sorokin, and L.~Tang, ``Metamaterial beam with graded
  local resonators for broadband vibration suppression,'' \emph{Mechanical
  Systems and Signal Processing}, vol. 146, p. 106982.

\bibitem{chen2020elastic}
Z.~Chen, Y.~Xia, J.~He, Y.~Xiong, and G.~Wang, ``Elastic-electro-mechanical
  modeling and analysis of piezoelectric metamaterial plate with a self-powered
  synchronized charge extraction circuit for vibration energy harvesting,''
  \emph{Mechanical Systems and Signal Processing}, vol. 143, p. 106824, 2020.

\bibitem{sirota2019tunable}
L.~Sirota, F.~Semperlotti, and A.~M. Annaswamy, ``Tunable and reconfigurable
  mechanical transmission-line metamaterials via direct active feedback
  control,'' \emph{Mechanical Systems and Signal Processing}, vol. 123, pp.
  117--130, 2019.

\bibitem{sirota2020modeling}
L.~Sirota and A.~M. Annaswamy, ``Active boundary and interior absorbers for
  one-dimensional wave propagation: {A}pplication to transmission-line
  metamaterials,'' \emph{Automatica}, vol. 117, pp. 108--855, 2020.

\bibitem{thouless1982quantized}
D.~J. Thouless, M.~Kohmoto, M.~P. Nightingale, and M.~den Nijs, ``Quantized
  {H}all conductance in a two-dimensional periodic potential,'' \emph{Physical
  Review Letters}, vol.~49, no.~6, p. 405, 1982.

\bibitem{haldane1988model}
F.~D.~M. Haldane, ``Model for a quantum {H}all effect without {L}andau levels:
  {C}ondensed-matter realization of the ``parity anomaly",'' \emph{Physical
  Review Letters}, vol.~61, no.~18, p. 2015, 1988.

\bibitem{kane2005quantum}
C.~L. Kane and E.~J. Mele, ``Quantum spin {H}all effect in graphene,''
  \emph{Physical Review Letters}, vol.~95, no.~22, p. 226801, 2005.

\bibitem{bernevig2006quantum}
B.~A. Bernevig, T.~L. Hughes, and S.-C. Zhang, ``Quantum spin {H}all effect and
  topological phase transition in hgte quantum wells,'' \emph{Science}, vol.
  314, no. 5806, pp. 1757--1761, 2006.

\bibitem{franz2013topological}
M.~Franz and L.~Molenkamp, \emph{Topological Insulators}.\hskip 1em plus 0.5em
  minus 0.4em\relax Elsevier, 2013.

\bibitem{lu2014topological}
L.~Lu, J.~D. Joannopoulos, and M.~Solja{\v{c}}i{\'c}, ``Topological
  photonics,'' \emph{Nature Photonics}, vol.~8, no.~11, p. 821, 2014.

\bibitem{rechtsman2013photonic}
M.~C. Rechtsman, J.~M. Zeuner, Y.~Plotnik, Y.~Lumer, D.~Podolsky, F.~Dreisow,
  S.~Nolte, M.~Segev, and A.~Szameit, ``Photonic floquet topological
  insulators,'' \emph{Nature}, vol. 496, no. 7444, p. 196, 2013.

\bibitem{peano2016topological}
V.~Peano, M.~Houde, C.~Brendel, F.~Marquardt, and A.~A. Clerk, ``Topological
  phase transitions and chiral inelastic transport induced by the squeezing of
  light,'' \emph{Nature Communications}, vol.~7, no.~1, pp. 1--8, 2016.

\bibitem{yang2015topological}
Z.~Yang, F.~Gao, X.~Shi, X.~Lin, Z.~Gao, Y.~Chong, and B.~Zhang, ``Topological
  acoustics,'' \emph{Physical Review Letters}, vol. 114, no.~11, p. 114301,
  2015.

\bibitem{zhang2017topological}
Z.~Zhang, Q.~Wei, Y.~Cheng, T.~Zhang, D.~Wu, and X.~Liu, ``Topological creation
  of acoustic pseudospin multipoles in a flow-free symmetry-broken metamaterial
  lattice,'' \emph{Physical Review Letters}, vol. 118, no.~8, p. 084303, 2017.

\bibitem{yves2017topological}
S.~Yves, R.~Fleury, F.~Lemoult, M.~Fink, and G.~Lerosey, ``Topological acoustic
  polaritons: robust sound manipulation at the subwavelength scale,'' \emph{New
  Journal of Physics}, vol.~19, no.~7, p. 075003, 2017.

\bibitem{sussman2016topological}
D.~M. Sussman, O.~Stenull, and T.~Lubensky, ``Topological boundary modes in
  jammed matter,'' \emph{Soft Matter}, vol.~12, no.~28, pp. 6079--6087, 2016.

\bibitem{pal2017edge}
R.~K. Pal and M.~Ruzzene, ``Edge waves in plates with resonators: an elastic
  analogue of the {Q}uantum {V}alley {H}all {E}ffect,'' \emph{New Journal of
  Physics}, vol.~19, no.~2, p. 025001, 2017.

\bibitem{chaunsali2018subwavelength}
R.~Chaunsali, C.-W. Chen, and J.~Yang, ``Subwavelength and directional control
  of flexural waves in zone-folding induced topological plates,''
  \emph{Physical Review B}, vol.~97, no.~5, p. 054307, 2018.

\bibitem{wang2015topological}
P.~Wang, L.~Lu, and K.~Bertoldi, ``Topological phononic crystals with one-way
  elastic edge waves,'' \emph{Physical Review Letters}, vol. 115, no.~10, p.
  104302, 2015.

\bibitem{nash2015topological}
L.~M. Nash, D.~Kleckner, A.~Read, V.~Vitelli, A.~M. Turner, and W.~T. Irvine,
  ``Topological mechanics of gyroscopic metamaterials,'' \emph{Proceedings of
  the National Academy of Sciences}, vol. 112, no.~47, pp. 14\,495--14\,500,
  2015.

\bibitem{pan2014valley}
H.~Pan, Z.~Li, C.-C. Liu, G.~Zhu, Z.~Qiao, and Y.~Yao, ``Valley-polarized
  quantum anomalous {H}all effect in silicene,'' \emph{Physical Review
  Letters}, vol. 112, no.~10, p. 106802, 2014.

\bibitem{zhou2018quantum}
Y.~Zhou, P.~R. Bandaru, and D.~F. Sievenpiper, ``Quantum-spin-{H}all
  topological insulator in a spring-mass system,'' \emph{New Journal of
  Physics}, vol.~20, no.~12, p. 123011, 2018.

\bibitem{susstrunk2015observation}
R.~S{\"u}sstrunk and S.~D. Huber, ``Observation of phononic helical edge states
  in a mechanical topological insulator,'' \emph{Science}, vol. 349, no. 6243,
  pp. 47--50, 2015.

\bibitem{zhang2018achieving}
Z.~Zhang, Y.~Cheng, and X.~Liu, ``Achieving acoustic topological valley-{H}all
  states by modulating the subwavelength honeycomb lattice,'' \emph{Scientific
  Reports}, vol.~8, no.~1, pp. 1--8, 2018.

\bibitem{zhou2020voltage}
W.~Zhou, Y.~Su, W.~Chen, C.~Lim \emph{et~al.}, ``Voltage-controlled quantum
  valley {H}all effect in dielectric membrane-type acoustic metamaterials,''
  \emph{International Journal of Mechanical Sciences}, vol. 172, p. 105368,
  2020.

\bibitem{darabi2020experimental}
A.~Darabi, M.~Collet, and M.~J. Leamy, ``Experimental realization of a
  reconfigurable electroacoustic topological insulator,'' \emph{Proceedings of
  the National Academy of Sciences}, vol. 117, no.~28, pp. 16\,138--16\,142,
  2020.

\bibitem{hofmann2019chiral}
T.~Hofmann, T.~Helbig, C.~H. Lee, M.~Greiter, and R.~Thomale, ``Chiral voltage
  propagation and calibration in a topolectrical {C}hern circuit,''
  \emph{Physical Review Letters}, vol. 122, no.~24, p. 247702, 2019.

\bibitem{scheibner2020non}
C.~Scheibner, W.~T. Irvine, and V.~Vitelli, ``Non-{H}ermitian band topology and
  skin modes in active elastic media,'' \emph{Physical {R}eview {L}etters},
  vol. 125, no.~11, p. 118001, 2020.

\bibitem{rosa2020dynamics}
M.~I. Rosa and M.~Ruzzene, ``Dynamics and topology of non-{H}ermitian elastic
  lattices with non-local feedback control interactions,'' \emph{New {J}ournal
  of {P}hysics}, vol.~22, no.~5, p. 053004, 2020.

\bibitem{brandenbourger2019non}
M.~Brandenbourger, X.~Locsin, E.~Lerner, and C.~Coulais, ``Non-reciprocal
  robotic metamaterials,'' \emph{Nature Communications}, vol.~10, no.~1, pp.
  1--8, 2019.

\bibitem{kotwal2019active}
T.~Kotwal, H.~Ronellenfitsch, F.~Moseley, and J.~Dunkel, ``Active topolectrical
  circuits,'' \emph{arXiv preprint arXiv:1903.10130}, 2019.

\bibitem{sirota2020feedback}
L.~Sirota, R.~Ilan, Y.~Shokef, and Y.~Lahini, ``Non-{N}ewtonian topological
  mechanical metamaterials using feedback control,'' \emph{arXiv preprint
  arXiv:2002.10607}, 2020.

\bibitem{sirota2020feedbackA}
L.~Sirota, Y.~Lahini, R.~Ilan, and Y.~Shokef, ``Feedback-based topological
  mechanical metamaterials,'' \emph{Accepted to the 14th International Congress
  on Artificial Materials for Novel Wave Phenomena}, 2020.

\bibitem{pierce1990acoustics}
A.~D. Pierce and R.~T. Beyer, ``Acoustics: {A}n introduction to its physical
  principles and applications. 1989 edition,'' 1990.

\bibitem{ginsberg2018acoustics}
J.~H. Ginsberg, \emph{Acoustics: A Textbook for Engineers and
  Physicists}.\hskip 1em plus 0.5em minus 0.4em\relax Springer, 2018, vol.~2.

\bibitem{curtain2009transfer}
R.~Curtain and K.~Morris, ``Transfer functions of distributed parameter
  systems: A tutorial,'' \emph{Automatica}, vol.~45, no.~5, pp. 1101--1116,
  2009.

\bibitem{becker2018immersive}
T.~S. Becker, D.-J. van Manen, C.~M. Donahue, C.~B{\"a}rlocher, N.~B{\"o}rsing,
  F.~Broggini, T.~Haag, J.~O. Robertsson, D.~R. Schmidt, S.~A. Greenhalgh
  \emph{et~al.}, ``Immersive wave propagation experimentation: Physical
  implementation and one-dimensional acoustic results,'' \emph{Physical Review
  X}, vol.~8, no.~3, p. 031011, 2018.

\bibitem{hu2019modelling}
G.~Hu, L.~Tang, and X.~Cui, ``On the modelling of membrane-coupled helmholtz
  resonator and its application in acoustic metamaterial system,''
  \emph{Mechanical Systems and Signal Processing}, vol. 132, pp. 595--608,
  2019.

\bibitem{xiao2012flexural}
Y.~Xiao, J.~Wen, and X.~Wen, ``Flexural wave band gaps in locally resonant thin
  plates with periodically attached spring--mass resonators,'' \emph{Journal of
  Physics D: Applied Physics}, vol.~45, no.~19, p. 195401, 2012.

\bibitem{wang2015tuning}
X.-P. Wang, P.~Jiang, T.-N. Chen, and J.~Zhu, ``Tuning characteristic of band
  gap and waveguide in a multi-stub locally resonant phononic crystal plate,''
  \emph{AIP Advances}, vol.~5, no.~10, p. 107141, 2015.

\bibitem{brillouin1953wave}
L.~Brillouin, ``Wave propagation in periodic structures: electric filters and
  crystal lattices,'' 1953.

\bibitem{reich2002tight}
S.~Reich, J.~Maultzsch, C.~Thomsen, and P.~Ordejon, ``Tight-binding description
  of graphene,'' \emph{Physical Review B}, vol.~66, no.~3, p. 035412, 2002.

\bibitem{sirota2019active}
L.~Sirota and A.~M. Annaswamy, ``Active wave suppression in the interior of a
  one-dimensional domain,'' \emph{Automatica}, vol. 100, pp. 403--406, 2019.

\bibitem{sirota2010free}
L.~Sirota and Y.~Halevi, ``Free response and absolute vibration suppression of
  second-order flexible structures—the traveling wave approach,''
  \emph{Journal of Vibration and Acoustics}, vol. 132, no.~3, p. 031008, 2010.

\bibitem{sirota2015fractional}
------, ``Fractional order control of flexible structures governed by the
  damped wave equation,'' in \emph{American Control Conference (ACC),
  2015}.\hskip 1em plus 0.5em minus 0.4em\relax IEEE, 2015, pp. 565--570.

\bibitem{sirota2015fractionalA}
------, ``Fractional order control of the two-dimensional wave equation,''
  \emph{Automatica}, vol.~59, pp. 152--163, 2015.

\bibitem{wilcox1977theory}
C.~H. Wilcox, ``Theory of bloch waves.'' {U}tah {U}niv {S}alt {L}ake {C}ity
  {D}ept of {M}athematics, Tech. Rep., 1977.

\bibitem{Fukui2005}
T.~Fukui, Y.~Hatsugai, and H.~Suzuki, ``Chern numbers in discretized brillouin
  zone: Efficient method of computing (spin) hall conductances,'' \emph{Journal
  of the Physical Society of Japan}, vol.~74, no.~6, pp. 1674--1677, 2005.

\end{thebibliography}
\end{document}